\documentclass[prd,twocolumn,showpacs,superscriptaddress,nofootinbib,floatfix,10pt]{revtex4-2}
\usepackage{bm,times,braket,amsfonts,amssymb,stmaryrd,latexsym,amsmath}
\usepackage[usenames,dvipsnames]{color}
\usepackage{epsfig,slashed,hyperref,graphicx,orcidlink,multirow,booktabs}
\allowdisplaybreaks[1]
\definecolor{nicered}{rgb}{0.7,0.1,0.1}
\definecolor{nicegreen}{rgb}{0.1,0.5,0.1}
\definecolor{navyblue}{RGB}{0,110,184}
\hypersetup{colorlinks,citecolor=nicegreen,linkcolor=nicered,urlcolor=navyblue}

\newcommand{\Lcb}{\bm{L}\!\cdot\!(\bm{S}_i\!+\!\bm{S}_j)}
\newcommand{\jl}{j_\ell}

\begin{document}

\title{The excited baryon spectrum from a unified quark model}

\author{Yan-Ke Chen \orcidlink{0000-0002-9984-163X}}
\affiliation{School of Physics and Center of High Energy Physics, Peking University, Beijing 100871, China}

\author{Liang-Zhen Wen\,\orcidlink{0009-0006-8266-5840}}\email{wenlzh\_hep-th@stu.pku.edu.cn} 
\affiliation{School of Physics and Center of High Energy Physics, Peking University, Beijing 100871, China}

\author{Wei-Lin Wu\,\orcidlink{0009-0009-3480-8810}}\email{wlwu@pku.edu.cn}
\affiliation{School of Physics, Peking University, Beijing 100871, China}

\author{Lu Meng \orcidlink{0000-0001-9791-7138}}\email{lmeng@seu.edu.cn}
\affiliation{School of Physics, Southeast University, Nanjing 211189, China}

\author{Shi-Lin Zhu\,\orcidlink{0000-0002-4055-6906}}\email{zhusl@pku.edu.cn}
\affiliation{School of Physics and Center of High Energy Physics, Peking University, Beijing 100871, China}

\begin{abstract}
A common approach to studying the multiquark states is to solve the few-body Schr\"odinger equation within a quark potential model. The multiquark states may contain quarks with several different flavors and have richer color structures than ordinary hadrons. The reliability of such an investigation requires that the underlying quark potential model can simultaneously describe the meson and baryon spectra across all flavor sectors, including orbital and radial excitations, with a single parameter set. At present, no quark potential model fully satisfies this requirement. We construct a simple nonrelativistic constituent quark potential model that includes spin--orbit and tensor interactions. We refit the model parameters to the latest experimental data. The excited light hadrons and several exotic hadron candidates are excluded from the fit. The resulting parameter set reproduces the spectra across all fitted sectors. The vast majority of deviations are below $20$ MeV, while the largest deviations remain of the order of several tens of MeV, which is the typical accuracy of nonrelativistic quark potential models. Using the same parameters, we predict the spectra and internal structures of the orbitally and radially excited heavy baryons, which await further experimental determination. The excited light baryons, deliberately left out of the fit, are calculated with the same parameters. The calculated excited light baryon spectra deviate substantially from experimental values. Our refitted model does not resolve these longstanding discrepancies. The results delimit the range of validity of the nonrelativistic constituent quark potential model. By treating mesons and baryons across all flavor sectors, including both orbital and radial excitations within a unified framework, the model provides a controlled starting point for few-body calculations of multiquark states. We also urge experimental searches for the predicted states.
\end{abstract}

\maketitle

\section{Introduction}\label{sec:intro}

Over the past two decades, the experimental discovery of numerous exotic hadrons that cannot be accommodated as conventional $q\bar q$ mesons or $qqq$ baryons has revitalized hadron spectroscopy~\cite{Chen:2016qju,Lebed:2016hpi,Chen:2016spr,Esposito:2016noz,Guo:2017jvc,Olsen:2017bmm,Brambilla:2019esw,Liu:2019zoy,Chen:2022asf,Meng:2022ozq,Liu:2024uxn,Bai:2026atm}. Recently, the number of reported multiquark candidate states has increased rapidly, including the fully-charmed $X(6900)$ in the di-$J/\psi$ spectrum~\cite{LHCb:2020bwg,CMS:2023owd,ATLAS:2023bft}, the open-charm--strange $T_{cs0}(2900)$ and the doubly-charged $T_{c\bar s}(2900)$~\cite{LHCb:2020bls,LHCb:2020pxc,LHCb:2022lzp,LHCb:2022sfr}, the charged $Z_c(3900)$ and its strange partner $Z_{cs}(3985)$~\cite{BESIII:2013ris,BESIII:2020qkh}, and the doubly-charmed $T_{cc}(3875)^+$~\cite{LHCb:2021vvq,LHCb:2021auc}.

Whether these states are compact multiquarks or hadronic molecules remains under active debate. To draw definitive conclusions, one requires a unified dynamical framework that can treat both pictures on an equal footing. The constituent quark model provides such a framework. By solving the few-body Schr\"odinger equation that admits both compact and well-separated cluster configurations, one can let the dynamical calculation itself decide the dominant picture~\cite{Ma:2022vqf,Meng:2023jqk,Chen:2023syh,Wu:2024zbx,Ma:2023int}. The multiquark state is more sensitive to the details of the interaction than a conventional hadron. The color structure of a multiquark state is richer: a $q\bar q$ pair may form a color $\bm 1_c$ or $\bm 8_c$, and a $qq$ pair may form a $\bar{\bm 3}_c$ or $\bm 6_c$. Furthermore, a single multiquark state may contain quarks with several different flavors. Therefore, a quark potential model that can reliably describe multiquark states should use a single parameter set to reproduce the spectra of mesons and baryons across all flavor sectors simultaneously. This ensures that each two- and three-body subsystem in the multiquark system is under control.

Most quark potential models, however, do not fully satisfy this requirement. The relativized Godfrey-Isgur (GI) model describes the mesons of all flavors in a single framework~\cite{Godfrey:1985xj}. The nonrelativistic potential model of Barnes, Godfrey, and Swanson describes the charmonium and bottomonium spectra~\cite{Barnes:2005pb,Godfrey:2015dia}. Capstick and Isgur extended the GI model to baryons, retaining the quark masses and one gluon exchange (OGE) parameters from the meson fit while introducing a separate string tension and additive constant for the baryon sector~\cite{Capstick:1986ter}. In addition, many heavy baryons had not yet been observed at the time. Only part of the original GI heavy baryon spectrum could be tested against data. More recently, the GI model was refitted to updated experimental data, and the excited singly heavy baryon spectrum was revisited systematically~\cite{Weng:2024roa}. The chiral constituent quark models, which supplement the OGE with Goldstone-boson exchanges between the light quarks, describe the meson and the heavy baryon spectra~\cite{Vijande:2004he,Segovia:2013wma,Valcarce:2008dr}, but the two sectors are constrained separately rather than fitted simultaneously. The relativistic quasipotential model of Ebert, Faustov, and Galkin covers the heavy-light mesons~\cite{Ebert:2009ua} and the heavy baryons~\cite{Ebert:2005xj,Ebert:2007nw,Ebert:2011kk} by replacing the genuine three-body baryon dynamics with an effective quark--diquark description. Other constituent quark models have been used to calculate orbital and radial excitations in individual light- and heavy-baryon sectors~\cite{Zhong:2024mnt,Liu:2019wdr,Chen:2016iyi,Liu:2019vtx,Zhou:2025fpp}. QCD sum rules also provide a complementary route to the spectra of excited hadrons~\cite{Wang:2010it,Wang:2020mxk,Chen:2015kpa,Mao:2015gya,Chen:2016phw,Chen:2017sbg}.

The line of work by Semay and Silvestre-Brac better satisfies the ``unified'' requirement. The nonrelativistic AL1/AP1 potentials are fitted with a single parameter set to the meson and baryon spectra of all flavors~\cite{Semay:1994ht,Silvestre-Brac:1996myf}, and a semirelativistic instanton-induced interaction provides a unified description of the light mesons and baryons~\cite{Brau:2002zpy}. Furthermore, the AL1 potential has a simple local form and is easy to incorporate into well-established solvers for few-body problems. Many few-body studies based on the AL1 model have successfully described multiquark states such as the $T_{cc}(3875)$, the $T_{cs0}(2900)$, and the $X(6900)$, and predicted further bound and resonant multiquark states awaiting experimental observation~\cite{Meng:2023jqk,Chen:2023syh,Ma:2023int,Wu:2024zbx,Wu:2024euj,Wu:2024hrv,Zheng:2025uzy}. However, the AL1 model still has two limitations for our purposes. First, it contains neither a spin--orbit nor a tensor force. It reproduces only the $S$-wave ground states, but cannot describe the orbitally or radially excited states which are of increasing interest in multiquark state research. Second, its parameters were fixed more than two decades ago, before many excited heavy baryons were observed.

The experimental progress in the heavy baryon sector has been substantial since the original AL1 fits. The negative-parity $\Lambda_c(2595)/\Lambda_c(2625)$, $\Xi_c(2790)/\Xi_c(2815)$, and $\Lambda_b(5912)/\Lambda_b(5920)$~\cite{LHCb:2012kxf} doublets are now firmly established. In the charm sector, the LHCb and Belle collaborations observed the five $\Omega_c(3000$--$3120)$ states~\cite{LHCb:2017uwr,Belle:2017ext}, the $\Lambda_c(2860)$ and $\Lambda_c(2910)$~\cite{LHCb:2017jym,Belle:2022hnm}, the $\Xi_c(2923/2939/2965)$~\cite{LHCb:2020iby}, and the $\Omega_c(3185/3227)$~\cite{LHCb:2023sxp}. In the bottom sector, the observed states include $\Sigma_b(6097)$~\cite{LHCb:2018haf}, the $\Xi_b(6227)$~\cite{LHCb:2018vuc}, the $\Lambda_b(6146/6152)$~\cite{LHCb:2019soc,CMS:2020zzv}, the $\Omega_b(6316$--$6350)$~\cite{LHCb:2020tqd}, the $\Xi_b(6100)$~\cite{CMS:2021rvl}, and further excited $\Xi_b$ states~\cite{LHCb:2021ssn}. Significant progress has also been achieved in the doubly heavy sector. An early $\Xi_{cc}^{+}$ signal was reported by SELEX~\cite{SELEX:2001fbx} but was not confirmed. The $\Xi_{cc}^{++}$ was observed by LHCb~\cite{LHCb:2017iph}. Most recently, its partner $\Xi_{cc}^{+}$ and $\Omega_{cc}$ have also been observed~\cite{LHCb:2026pxn,LHCb:2026omegacc}. Therefore, an updated global fit is timely and warranted. These measurements provide a substantially broader basis for a global fit.

In this work, we address both limitations above. We add the spin--orbit term and the tensor term to the AL1 potential. We refit the model parameters globally to the up-to-date experimental data set comprising the $1S$, $2S$, $1P$ mesons and the $1S$, $1P$ baryons across all flavors. The excited  light hadrons and some typical exotic hadron candidates, such as the Roper resonance $N(1440)$, $N(1535)$, $\Lambda(1405)$, $a_0(980)$, $f_0(980)$, $D_{s0}^*(2317)$ and $D_{s1}(2460)$, still cannot be satisfactorily described by present conventional quark potential models~\cite{Guo:2017jvc,Chen:2022asf,Meng:2022ozq}. We exclude them from the global fit. 

The resulting single parameter set reproduces the entire data set to an accuracy of tens of MeV. With these parameters, we predict the unobserved doubly- and triply-heavy $1S$ baryons. We also determine the spectra, compositions, and sizes of excited heavy baryons. Finally, we calculate the excited light baryon spectrum, which is not included in the fit. The calculated excited light baryon spectrum deviates significantly from experimental results, indicating that the refit does not resolve the long-standing problem. This sector delimits the range of validity of the model. The model treats mesons and baryons across all flavor sectors with a unified framework and is intended to provide a foundation for subsequent few-body calculations of multiquark states.

The paper is organized as follows. In Sec.~\ref{sec:formalism}, we introduce the Hamiltonian and the Gaussian expansion method, and describe the global fit. Section~\ref{sec:fit_res} presents the fitted parameters and the resulting meson and baryon spectra. Section~\ref{sec:result} presents the predictions obtained with the same parameter set. A brief summary is given in Sec.~\ref{sec:summary}.

\section{Formalism}\label{sec:formalism}

\subsection{Hamiltonian}\label{sec:ham}
In the nonrelativistic constituent quark model, the Hamiltonian of an $N$-body system in the center-of-mass frame reads
\begin{equation}\label{eq:H}
  H=\sum_{i}\Big(m_i+\frac{p_i^2}{2 m_i}\Big)-T_{\mathrm{c.m.}}+\sum_{i<j} V_{ij}+V_{3b},
\end{equation}
where $m_i$ and $p_i$ are the mass and momentum of the $i$th (anti)quark. $T_{\mathrm{c.m.}}$ is the center-of-mass kinetic energy. We adopt the AL1 quark potential~\cite{Semay:1994ht,Silvestre-Brac:1996myf} and add the spin--orbit and tensor interactions. The two-body potential is
\begin{equation}\label{eq:Vij}
  V_{ij}=-\tfrac{3}{16}\,\bm{\lambda}^c_i\!\cdot\!\bm{\lambda}^c_j
  \big(V_{\mathrm{cen}}+V_{\mathrm{so}}+V_{\mathrm{ts}}\big),
\end{equation}
where $\bm{\lambda}^c_i$ are the color Gell-Mann matrices. The central, spin--orbit, and tensor interactions are
\begin{align}
  V_{\mathrm{cen}}={}&-\frac{4 \alpha_s}{3 r}+\lambda r-\Lambda \nonumber\\
    &+\frac{32 \pi \alpha_s g_{ss}}{9\,m_i m_j}\,\delta^{(3)}(\bm r)\,\bm{S}_i\!\cdot\!\bm{S}_j,
    \label{eq:Vcen}\\
  V_{\mathrm{so}}={}&\Lcb\Big[g_{\mathrm{ls},\alpha}\Big(\tfrac{1}{m_i^2}+\tfrac{1}{m_j^2}
    +\tfrac{4}{m_i m_j}\Big)\tfrac{\alpha_s}{3 r^3} \nonumber\\
    &\quad-g_{\mathrm{ls},\lambda}\Big(\tfrac{1}{m_i^2}+\tfrac{1}{m_j^2}\Big)\tfrac{\lambda}{4 r}\Big]
    \nonumber\\
    &+\bm{L}\!\cdot\!(\bm{S}_i\!-\!\bm{S}_j)\Big(\tfrac{1}{m_i^2}-\tfrac{1}{m_j^2}\Big) \nonumber\\
    &\quad\times\Big[g_{\mathrm{ls},\alpha}\tfrac{\alpha_s}{3 r^3}
    -g_{\mathrm{ls},\lambda}\tfrac{\lambda}{4 r}\Big],
    \label{eq:Vso}\\
  V_{\mathrm{ts}}={}&\frac{4 \alpha_s g_{\mathrm{tensor}}}{3\,m_i m_j\,r^3} \nonumber\\
    &\times\Big[\frac{3(\bm{S}_i\!\cdot\!\bm r)(\bm{S}_j\!\cdot\!\bm r)}{r^2}
    -\bm{S}_i\!\cdot\!\bm{S}_j\Big],
    \label{eq:Vts}
\end{align}
respectively. The first bracket of $V_{\mathrm{so}}$ is the OGE spin--orbit force, and the second is the Thomas-precession term induced by the scalar linear confinement. We treat $g_{\mathrm{ls},\alpha}$ and $g_{\mathrm{ls},\lambda}$ as independent parameters, which improves the description of the orbitally excited spectra. The $\bm{L}\!\cdot\!(\bm{S}_i-\bm{S}_j)$ part is antisymmetric in the two spins and is active only for unequal masses. The $\delta^{(3)}(\bm r)$ and $1/r^3$ operators are short-range artifacts of the nonrelativistic reduction and are regularized as
\begin{equation}\label{eq:reg}
  \delta^{(3)}(\bm r)\to\frac{e^{-r^2/r_0^2}}{\pi^{3/2}r_0^3},\qquad
  \frac{1}{r^3}\to\frac{1-e^{-r^2/r_0^2}}{r^3},
\end{equation}
with a flavor-dependent range 
\begin{equation}
  r_0(m_i,m_j)=A\left(\frac{2 m_i m_j}{m_i+m_j}\right)^{-B}.
\end{equation}
The calculated spectra are insensitive to the $1/r^3$ regulator, so we use the same $r_0$ for $\delta^{(3)}(\bm r)$ and $1/r^3$ to reduce the number of parameters. The AL1 model also includes a three-body force,
\begin{equation}\label{eq:V3b}
  V_{3b}=\frac{C_{3b}}{m_1 m_2 m_3},
\end{equation}
with $C_{3b}=-2.02\times10^{-3}~\mathrm{GeV}^4$ in the original parametrization~\cite{Silvestre-Brac:1996myf}. 

The spin-orbit and tensor terms of Eqs.~(\ref{eq:Vso}) and~(\ref{eq:Vts}) are the only interactions we add to the original model. The free parameters are the four constituent masses $\{m_n,m_s,m_c,m_b\}$, the central parameters $\{\alpha_s,g_{ss},\lambda,\Lambda,A,B\}$, the spin-dependent strengths $\{g_{\mathrm{ls},\alpha},g_{\mathrm{ls},\lambda},g_{\mathrm{tensor}}\}$, and the three-body coupling $C_{3b}$.

\subsection{Gaussian expansion method}\label{sec:gem}

\begin{figure}[htbp]
  \centering
  \includegraphics[width=0.98\columnwidth]{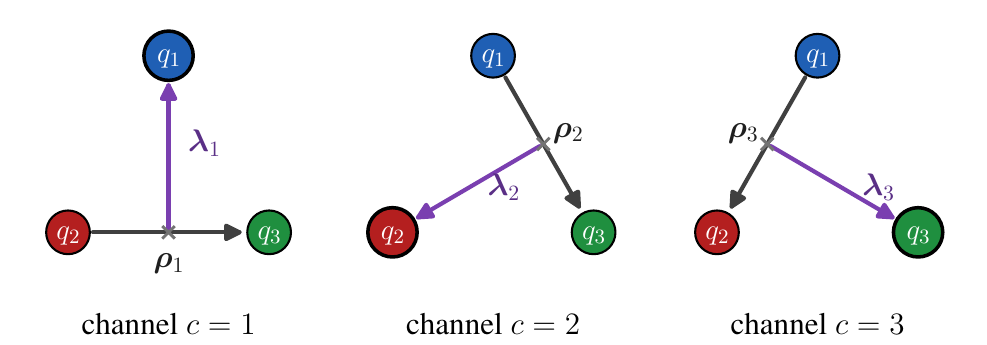}
  \caption{The three Jacobi coordinate channels for a three-body system. In channel $c$,
  the spectator quark is connected to the center of mass of the remaining pair by
  $\bm{\lambda}_c$, and the pair is described by its internal coordinate $\bm{\rho}_c$. The baryon wave function is expanded in all three channels.}
  \label{fig:jacobi}
\end{figure}

We calculate the two-body mesons and the three-body baryons with the Gaussian expansion method~\cite{Hiyama:2003cu}, which expands the wave function in a basis of Gaussian functions. For an $N$-body system, the internal motion is described by $N-1$ Jacobi relative coordinates. For a meson, there is a single $q\bar q$ separation. For a baryon, the three-body wave function is expanded in three Jacobi channels, as shown in Fig.~\ref{fig:jacobi}. The full wave function is
\begin{equation}\label{eq:psi}
  \Psi=\mathcal{A}\sum_{c}\sum_{\{n\}}C^{(c)}_{\{n\}}
  \Big[\prod_{k}\phi^{\mathrm G}_{n_k l_k m_k}\!\big(\bm{x}^{(c)}_k\big)\Big]
  \otimes\chi^{(c)}_{\mathrm{SCF}},
\end{equation}
where $c$ labels three Jacobi channels, $C^{(c)}_{\{n\}}$ are the undetermined expansion coefficients, $\chi^{(c)}_{\mathrm{SCF}}$ is the spin--color--flavor wave function, $\mathcal{A}$ antisymmetrizes the identical quarks, and each Gaussian function takes the form 
\begin{equation}
  \phi^{\mathrm G}_{nlm}(\bm r)=N_{nl}\,r^{l}e^{-\frac{r^2}{r_n^2}}Y_{lm}(\hat{\bm r}),
\end{equation}
where $N_{nl}$ is the normalization coefficient. $r_n$ is taken in geometric progression. $r_n=r_1 a^{n-1}$. $Y_{l m}$ is the spherical harmonic representing the angular momentum. Inserting Eq.~(\ref{eq:psi}) into the Schr\"odinger equation yields a generalized matrix eigenvalue problem. 

\subsection{Global fit}\label{sec:fit}
The parameters are determined by minimizing
\begin{equation}\label{eq:chi2}
  \chi^2=\sum_i\Big(\frac{M_{i,\mathrm{th}}-M_{i,\mathrm{exp}}}{\sigma_i}\Big)^2.
\end{equation}
For the hadron spectrum, the experimental uncertainties are far smaller than the intrinsic accuracy of the nonrelativistic quark model. Thus we choose $\sigma_i$ to be $10$--$20$~MeV.  We use a smaller $\sigma=10$ MeV for the fully heavy systems ($c\bar c,c\bar b,b\bar b$), where the nonrelativistic description is most reliable, and a larger $\sigma=20$ MeV for the lighter states.

The data set comprises the well-established hadrons whose interpretation as simple $q\bar q$ or $qqq$ states is uncontroversial. In the meson sector, it contains the $1S$ ground states of all flavors, the $2S$ radially excited charmonium, bottomonium, and $B_c$ mesons, and the $1P$ orbitally excited charmonium, bottomonium, and several heavy-light systems. In the baryon sector, it contains the $24$ $1S$ ground state baryons and $7$ $1P$ orbitally excited heavy baryons [$\Lambda_c(2595)$, $\Lambda_c(2625)$, $\Xi_c(2790)$, $\Xi_c(2815)$, $\Lambda_b(5912)$, $\Lambda_b(5920)$, $\Xi_b(6100)$].

Several states are deliberately excluded from the fit. The excited light mesons and baryons---for example $N(1440)$, $N(1535)$, $\Lambda(1405)$, $a_0(980)$, and $f_0(980)$---may be affected by strong coupled-channel effects and meson-cloud dynamics~\cite{Guo:2017jvc,Meng:2022ozq} that a conventional quark model cannot capture. We therefore leave the excited light hadrons out of the fit. The excited light baryons are nevertheless computed with the refitted parameters and are discussed in Sec.~\ref{sec:exc_light}. 

Among the heavy-light $1^+$ mesons, only the narrow $\jl=\tfrac32$ states are retained in the fitting process. In the heavy-quark limit, the two $J=1$ states are eigenstates of the light angular momentum $\bm{\jl}=\bm{l}+\bm{s}_q$, forming the narrow $\jl=\tfrac32$ and the broad $\jl=\tfrac12$ doublets~\cite{Isgur:1991wq}. The $\jl$ basis is related to the $^1\!P_1$--$^3\!P_1$ basis by a unitary rotation
\begin{equation}\label{eq:jl}
\begin{aligned}
  |\jl{=}\tfrac32\rangle &= \sqrt{\tfrac23}\,|{}^1\!P_1\rangle + \sqrt{\tfrac13}\,|{}^3\!P_1\rangle,\\
  |\jl{=}\tfrac12\rangle &= \sqrt{\tfrac13}\,|{}^1\!P_1\rangle - \sqrt{\tfrac23}\,|{}^3\!P_1\rangle.
\end{aligned}
\end{equation}
For each $J=1$ eigenstate $|\psi_k\rangle$, we calculate its $\jl=\tfrac32$ weight $P_{3/2,k}=|\langle\jl{=}\tfrac32|\psi_k\rangle|^2$. The narrow physical states [$D_1(2420)$, $D_{s1}(2536)$, $B_{s1}(5830)$] are eigenstates having large $P_{3/2,k}$ values. These states are included in the fit. Their broad $\jl=\tfrac12$ partners lie close to the $S$-wave thresholds and are strongly affected by the coupled-channel effects. Therefore, the scalar and axial states $D_0^*(2300)$, $D_{s0}^*(2317)$, $D_1(2430)$, and $D_{s1}(2460)$, which are widely interpreted as having large coupled-channel or molecular components~\cite{Guo:2017jvc,Chen:2022asf,Meng:2022ozq}, are neither fitted nor computed.

\section{Fit results}\label{sec:fit_res}
The refitted parameters are listed in Table~\ref{tab:params}, alongside the original AL1 values. Our new global fit reaches $\chi^2/\mathrm{dof}=1.04$. The $\chi^2$ contributions from the meson sector, the ground state baryon sector, and the excited baryon sector are approximately $42$, $6$, and $8$, respectively. The meson and baryon spectra are collected in Tables~\ref{tab:meson} and~\ref{tab:baryon}, respectively. The tables list theoretical masses and their deviations from experimental values $\Delta=M_{\mathrm{th}}-M_{\mathrm{exp}}$. The vast majority of deviations are below $20$ MeV, and the largest deviations are of the order of several tens of MeV, consistent with the intrinsic accuracy of the nonrelativistic quark model.

\begin{table}[htbp]\centering\small
\caption{Model parameters. ``AL1'' is the original set of Ref.~\cite{Silvestre-Brac:1996myf}. The notation ``---'' denotes
couplings absent from AL1.}
\label{tab:params}
\begin{tabular*}{\columnwidth}{@{\extracolsep{\fill}}llrr@{}}
\hline\hline
Parameter & Unit & AL1 & This work \\
\hline
$m_n$ & GeV & 0.315 & 0.3637 \\
$m_s$ & GeV & 0.577 & 0.6297 \\
$m_c$ & GeV & 1.836 & 1.8891 \\
$m_b$ & GeV & 5.227 & 5.2610 \\
$a_s$ &  & 0.3802 & 0.3625 \\
$g_{ss}$ &  & 3.6711 & 3.3478 \\
$\lambda$ & GeV$^2$ & 0.1653 & 0.1872 \\
$\Lambda$ & GeV & 0.8321 & 0.9677 \\
$A$ & GeV$^{B-1}$ & 1.6553 & 1.3411 \\
$B$ &  & 0.2204 & 0.2667 \\
$g_{\mathrm{ls},\alpha}$ &  & --- & 2.1730 \\
$g_{\mathrm{ls},\lambda}$ &  & --- & 0.9229 \\
$g_{\mathrm{tensor}}$ &  & --- & 2.1552 \\
$C_{3b}$ & GeV$^4$ & $-2.02\times10^{-3}$ & $-3.5154\times10^{-3}$ \\
\hline\hline
\end{tabular*}
\end{table}

As listed in Table~\ref{tab:params}, the spin--orbit and tensor strengths are now determined. The scalar-confinement coupling $g_{\mathrm{ls},\lambda}\approx0.9$ is comparable to its OGE counterpart $g_{\mathrm{ls},\alpha}$, indicating that both mechanisms are needed to reproduce the fine structure. The new three-body coupling is $C_{3b}=-3.5\times10^{-3}~\mathrm{GeV}^4$, about $1.7$ times the original AL1 value. Its $1/(m_1m_2m_3)$ scaling makes it most important for the light baryons, providing additional binding beyond the pairwise two-body potential.

\begin{table}[htbp]\centering\small
\caption{Meson spectrum (MeV). For each state the model mass $M$ and its deviation
$\Delta=M-M_{\rm exp}$ are listed in adjacent columns. ``AL1'' is the original parameter set of Ref.~\cite{Silvestre-Brac:1996myf} (no spin--orbit or tensor interaction, which leaves the $^{2S+1}P_J$ multiplets unsplit). The heavy-light $1^+$ states are the narrow $j_\ell=3/2$ levels obtained from the $^1P_1$--$^3P_1$ mixing. The experimental
masses are taken from Ref.~\cite{ParticleDataGroup:2024cfk}.}
\label{tab:meson}
\begin{tabular*}{\columnwidth}{@{\extracolsep{\fill}}lcrrrrr@{}}
\hline\hline
Meson & $J^P$ & $M_{\rm exp}$ & \multicolumn{2}{c}{AL1} & \multicolumn{2}{c}{This work} \\
\cmidrule(lr){4-5}\cmidrule(lr){6-7}
 & & & $M$ & $\Delta$ & $M$ & $\Delta$ \\
\hline
$\pi$ & $0^-$ & 138 & 138 & $+0$ & 136 & $-2$ \\
$\rho$ & $1^-$ & 775 & 770 & $-5$ & 755 & $-20$ \\
$\omega$ & $1^-$ & 783 & 770 & $-13$ & 755 & $-28$ \\
$K$ & $0^-$ & 496 & 491 & $-5$ & 474 & $-22$ \\
$K^*$ & $1^-$ & 894 & 904 & $+10$ & 902 & $+8$ \\
$\phi$ & $1^-$ & 1019 & 1021 & $+1$ & 1034 & $+15$ \\
$D$ & $0^-$ & 1867 & 1862 & $-5$ & 1843 & $-24$ \\
$D^*$ & $1^-$ & 2009 & 2016 & $+8$ & 2014 & $+5$ \\
$D_s$ & $0^-$ & 1968 & 1962 & $-6$ & 1955 & $-13$ \\
$D_s^*$ & $1^-$ & 2112 & 2102 & $-10$ & 2117 & $+4$ \\
$\eta_c$ & $0^-$ & 2984 & 3005 & $+21$ & 3002 & $+18$ \\
$J/\psi$ & $1^-$ & 3097 & 3101 & $+4$ & 3123 & $+27$ \\
$B$ & $0^-$ & 5280 & 5293 & $+14$ & 5258 & $-21$ \\
$B^*$ & $1^-$ & 5325 & 5350 & $+26$ & 5323 & $-2$ \\
$B_s$ & $0^-$ & 5367 & 5361 & $-6$ & 5340 & $-27$ \\
$B_s^*$ & $1^-$ & 5415 & 5417 & $+2$ & 5407 & $-8$ \\
$B_c$ & $0^-$ & 6274 & 6292 & $+17$ & 6275 & $+1$ \\
$\eta_b$ & $0^-$ & 9399 & 9424 & $+25$ & 9395 & $-4$ \\
$\Upsilon$ & $1^-$ & 9460 & 9461 & $+1$ & 9452 & $-8$ \\
\multicolumn{7}{l}{\emph{2S}}\\
$\eta_c(2S)$ & $0^-$ & 3638 & 3608 & $-30$ & 3646 & $+8$ \\
$\psi(2S)$ & $1^-$ & 3686 & 3641 & $-45$ & 3692 & $+6$ \\
$B_c(2S)$ & $0^-$ & 6871 & 6854 & $-17$ & 6870 & $-1$ \\
$\eta_b(2S)$ & $0^-$ & 9999 & 10003 & $+4$ & 9996 & $-3$ \\
$\Upsilon(2S)$ & $1^-$ & 10023 & 10012 & $-11$ & 10010 & $-13$ \\
\multicolumn{7}{l}{\emph{1P}}\\
$D_1(2420)$ & $1^+$ & 2422 & 2421 & $-1$ & 2409 & $-13$ \\
$D_2^*$ & $2^+$ & 2461 & 2452 & $-9$ & 2458 & $-3$ \\
$D_{s1}(2536)$ & $1^+$ & 2535 & 2473 & $-62$ & 2505 & $-30$ \\
$D_{s2}^*$ & $2^+$ & 2569 & 2504 & $-66$ & 2555 & $-15$ \\
$h_c$ & $1^+$ & 3525 & 3463 & $-63$ & 3498 & $-27$ \\
$\chi_{c0}$ & $0^+$ & 3415 & 3487 & $+72$ & 3412 & $-3$ \\
$\chi_{c1}$ & $1^+$ & 3511 & 3487 & $-24$ & 3503 & $-8$ \\
$\chi_{c2}$ & $2^+$ & 3556 & 3487 & $-70$ & 3544 & $-12$ \\
$B_2^*$ & $2^+$ & 5737 & 5794 & $+56$ & 5761 & $+24$ \\
$B_{s1}(5830)$ & $1^+$ & 5829 & 5811 & $-18$ & 5816 & $-12$ \\
$B_{s2}^*$ & $2^+$ & 5840 & 5824 & $-16$ & 5839 & $-1$ \\
$h_b$ & $1^+$ & 9899 & 9915 & $+16$ & 9900 & $+1$ \\
$\chi_{b0}$ & $0^+$ & 9859 & 9926 & $+67$ & 9863 & $+3$ \\
$\chi_{b1}$ & $1^+$ & 9893 & 9926 & $+33$ & 9902 & $+9$ \\
$\chi_{b2}$ & $2^+$ & 9912 & 9926 & $+14$ & 9926 & $+14$ \\
\hline\hline
\end{tabular*}
\end{table}

\begin{table}[htbp]\centering\small
\caption{Baryon spectrum (MeV). The theoretical mass $M$ and deviation $\Delta=M-M_{\rm exp}$ are listed in adjacent columns. The 24 ground states and the 7 heavy negative-parity ($1P$) baryons are included in the fit. ``AL1'' uses the original parameters of
Ref.~\cite{Silvestre-Brac:1996myf}. AL1 cannot split the $1P$ doublets, so its $\tfrac12^-$ and $\tfrac32^-$ partners are degenerate. The experimental masses are taken from Ref.~\cite{ParticleDataGroup:2024cfk}.}
\label{tab:baryon}
\begin{tabular*}{\columnwidth}{@{\extracolsep{\fill}}lcrrrrr@{}}
\hline\hline
Baryon & $J^P$ & $M_{\rm exp}$ & \multicolumn{2}{c}{AL1} & \multicolumn{2}{c}{This work} \\
\cmidrule(lr){4-5}\cmidrule(lr){6-7}
 & & & $M$ & $\Delta$ & $M$ & $\Delta$ \\
\hline
$N$ & $\tfrac12^+$ & 939 & 930 & $-9$ & 926 & $-13$ \\
$\Lambda$ & $\tfrac12^+$ & 1116 & 1115 & $-1$ & 1120 & $+4$ \\
$\Sigma$ & $\tfrac12^+$ & 1193 & 1194 & $+1$ & 1187 & $-6$ \\
$\Xi$ & $\tfrac12^+$ & 1318 & 1321 & $+3$ & 1330 & $+12$ \\
$\Delta$ & $\tfrac32^+$ & 1232 & 1242 & $+10$ & 1219 & $-13$ \\
$\Sigma^*$ & $\tfrac32^+$ & 1385 & 1402 & $+17$ & 1393 & $+8$ \\
$\Xi^*$ & $\tfrac32^+$ & 1533 & 1540 & $+7$ & 1547 & $+14$ \\
$\Omega$ & $\tfrac32^+$ & 1672 & 1664 & $-8$ & 1687 & $+15$ \\
$\Lambda_c$ & $\tfrac12^+$ & 2286 & 2279 & $-7$ & 2284 & $-2$ \\
$\Sigma_c$ & $\tfrac12^+$ & 2454 & 2454 & $-0$ & 2442 & $-12$ \\
$\Sigma_c^*$ & $\tfrac32^+$ & 2518 & 2535 & $+17$ & 2528 & $+10$ \\
$\Xi_c$ & $\tfrac12^+$ & 2469 & 2463 & $-6$ & 2472 & $+3$ \\
$\Xi_c'$ & $\tfrac12^+$ & 2579 & 2567 & $-12$ & 2570 & $-9$ \\
$\Xi_c^*$ & $\tfrac32^+$ & 2646 & 2645 & $-1$ & 2654 & $+8$ \\
$\Omega_c$ & $\tfrac12^+$ & 2695 & 2674 & $-21$ & 2692 & $-3$ \\
$\Omega_c^*$ & $\tfrac32^+$ & 2766 & 2749 & $-17$ & 2774 & $+8$ \\
$\Xi_{cc}$ & $\tfrac12^+$ & 3621 & 3606 & $-15$ & 3612 & $-9$ \\
$\Lambda_b$ & $\tfrac12^+$ & 5620 & 5631 & $+11$ & 5613 & $-7$ \\
$\Sigma_b$ & $\tfrac12^+$ & 5813 & 5844 & $+31$ & 5810 & $-3$ \\
$\Sigma_b^*$ & $\tfrac32^+$ & 5833 & 5874 & $+41$ & 5844 & $+11$ \\
$\Xi_b$ & $\tfrac12^+$ & 5794 & 5802 & $+8$ & 5787 & $-7$ \\
$\Xi_b'$ & $\tfrac12^+$ & 5935 & 5940 & $+5$ & 5922 & $-13$ \\
$\Xi_b^*$ & $\tfrac32^+$ & 5954 & 5971 & $+17$ & 5956 & $+2$ \\
$\Omega_b$ & $\tfrac12^+$ & 6046 & 6032 & $-14$ & 6029 & $-17$ \\
\hline \multicolumn{7}{l}{\emph{$1P$ (negative-parity, fitted)}}\\
$\Lambda_c(2595)$ & $\tfrac12^-$ & 2592 & 2623 & $+30$ & 2620 & $+27$ \\
$\Lambda_c(2625)$ & $\tfrac32^-$ & 2628 & 2623 & $-6$ & 2651 & $+23$ \\
$\Xi_c(2790)$ & $\tfrac12^-$ & 2792 & 2784 & $-8$ & 2786 & $-6$ \\
$\Xi_c(2815)$ & $\tfrac32^-$ & 2820 & 2784 & $-36$ & 2819 & $-1$ \\
$\Lambda_b(5912)$ & $\tfrac12^-$ & 5912 & 5957 & $+45$ & 5942 & $+30$ \\
$\Lambda_b(5920)$ & $\tfrac32^-$ & 5920 & 5957 & $+37$ & 5954 & $+34$ \\
$\Xi_b(6100)$ & $\tfrac32^-$ & 6100 & 6102 & $+3$ & 6106 & $+6$ \\
\hline\hline
\end{tabular*}
\end{table}

The comparison with AL1 is instructive. Without spin-orbit and tensor interactions, AL1 leaves the $^{2S+1}P_J$ multiplets degenerate. Therefore, its $1P$ deviations reach $\pm70$~MeV in the charmonium and bottomonium sectors. After we add the spin--orbit and tensor terms and refit the model, the deviations across the entire $1P$ sector are reduced to within a few tens of MeV. Furthermore, the correct energy-level order is obtained, including the $\chi_{cJ}$ and $\chi_{bJ}$ splittings and the $h_c,h_b$ positions. The three fitted heavy-light $1^+$ mesons are nearly pure $\jl=\tfrac32$. The $D_1(2420)$, $D_{s1}(2536)$, and $B_{s1}(5830)$ eigenstates have $0.997$, $0.92$, and $0.88$ $\jl=\tfrac32$ weights, respectively, justifying their identification as the narrow physical states. In the baryon sector, all $24$ ground states are reproduced to within $20$ MeV, and the seven excited heavy baryons are reproduced to within approximately $30$ MeV.

It should be stressed that many specialized models~\cite{Godfrey:1985xj,Barnes:2005pb,Godfrey:2015dia,Capstick:1986ter,Weng:2024roa,Vijande:2004he,Segovia:2013wma,Valcarce:2008dr,Ebert:2009ua,Ebert:2005xj,Ebert:2007nw,Ebert:2011kk,Semay:1994ht,Silvestre-Brac:1996myf,Brau:2002zpy} describe their respective sectors considerably better than our present model. However, our motivation is different. Our aim is to describe mesons and baryons across all flavor sectors, including both ground and excited states, within a unified framework and a single parameter set. This is a prerequisite for reliable multiquark few-body calculations. The $\sigma_i$ are fixed before minimization rather than tuned to enforce $\chi^2/\mathrm{dof}\simeq1$. The minimum near unity also reflects the practical accuracy limit of the present Hamiltonian with a single parameter set.

\section{Model predictions}\label{sec:result}
With the fixed parameters in Table~\ref{tab:params}, we apply the same Hamiltonian to three classes of systems. We predict the unobserved doubly- and triply-heavy $1S$ baryons. We further determine the spectra, compositions, and sizes of excited heavy baryons. Finally, we calculate the excited light baryon spectrum. No model parameter is refitted in the calculations below, and no state discussed here enters the fit except the seven $1P$ baryons already listed in Table~\ref{tab:baryon}.

\subsection{Ground states of the doubly- and triply-heavy baryons}\label{sec:pred}
Many heavy baryons have not yet been observed in experiments. The present model is in a good position to make predictions. The predicted doubly- and triply-heavy ground states are collected in Table~\ref{tab:pred}. For the doubly-charmed sector, the $\Xi_{cc}^{++}$~\cite{LHCb:2017iph} and its isospin partner $\Xi_{cc}^{+}$~\cite{LHCb:2026pxn} are observed by LHCb and enter the fit. Very recently, the $\Omega_{cc}$ was observed as well, with a mass of approximately 3727 MeV announced by LHCb~\cite{LHCb:2026omegacc}. It does not enter our fit data set, while its measured mass agrees with the predicted value of $3728$~MeV to within about $1$~MeV. The remaining doubly-heavy $bb$ and $bc$ baryons and the triply-heavy $\Omega_{ccc},\Omega_{ccb},\Omega_{bbc},\Omega_{bbb}$ await further experimental observation.

The orbital ($1P$) and radial ($2S$) excitations of both these systems and the singly heavy systems are discussed in Sec.~\ref{sec:exc}.

\begin{table}[htbp]\centering\small
\caption{Predicted ground-state masses (MeV) of the doubly- and triply-heavy baryons.
Spin-$\tfrac32$ states are starred. The $\Xi_{cc}^{++}$~\cite{LHCb:2017iph} and $\Xi_{cc}^{+}$~\cite{LHCb:2026pxn} are
established and enter the fit (Table~\ref{tab:baryon}). The $\Omega_{cc}$, observed near
$3727$~MeV~\cite{LHCb:2026omegacc}, does not enter the fit data set.}
\label{tab:pred}
\begin{tabular*}{\columnwidth}{@{\extracolsep{\fill}}lcrlcr@{}}
\hline\hline
Baryon & $J^P$ & $M$ & Baryon & $J^P$ & $M$ \\
\hline
$\Xi_{cc}^{*}$ & $\tfrac32^+$ & 3713 & $\Xi_{bc}$ & $\tfrac12^+$ & 6897 \\
$\Omega_{cc}$ & $\tfrac12^+$ & 3728 & $\Xi_{bc}'$ & $\tfrac12^+$ & 6944 \\
$\Omega_{cc}^{*}$ & $\tfrac32^+$ & 3819 & $\Xi_{bc}^{*}$ & $\tfrac32^+$ & 6979 \\
$\Xi_{bb}$ & $\tfrac12^+$ & 10157 & $\Omega_{bc}$ & $\tfrac12^+$ & 6999 \\
$\Xi_{bb}^{*}$ & $\tfrac32^+$ & 10200 & $\Omega_{bc}'$ & $\tfrac12^+$ & 7040 \\
$\Omega_{bb}$ & $\tfrac12^+$ & 10245 & $\Omega_{bc}^{*}$ & $\tfrac32^+$ & 7076 \\
$\Omega_{bb}^{*}$ & $\tfrac32^+$ & 10288 &  & &  \\
\hline
$\Omega_{ccc}$ & $\tfrac32^+$ & 4833 &  & &  \\
$\Omega_{ccb}$ & $\tfrac12^+$ & 8023 &  & &  \\
$\Omega_{ccb}^{*}$ & $\tfrac32^+$ & 8059 &  & &  \\
$\Omega_{bbc}$ & $\tfrac12^+$ & 11200 &  & &  \\
$\Omega_{bbc}^{*}$ & $\tfrac32^+$ & 11241 &  & &  \\
$\Omega_{bbb}$ & $\tfrac32^+$ & 14377 &  & &  \\
\hline\hline
\end{tabular*}
\end{table}

\subsection{Excited heavy baryons}\label{sec:exc}
We calculate the orbitally ($1P$) and radially ($2S$) excited heavy baryon spectra. The light baryons are treated separately in Sec.~\ref{sec:exc_light}. The $1P$ multiplets are obtained from a coupled $S=\tfrac12\oplus\tfrac32$ diagonalization that includes the $V_{\mathrm{so}}+V_{\mathrm{ts}}$ fine structure. The $2S$ states are the first radial excitations of the $L=0$ ground states, where the spin--orbit and tensor interactions vanish.

\subsubsection{Spectrum}\label{sec:exc_spec}
The $1P$ eigenvalues, classified by flavor symmetry, are listed in Appendix~\ref{app:spectra} (Tables~\ref{tab:1P_Qnn}--\ref{tab:1P_Q1Q1Q2}). For each baryon and each $J^P$, we list all states of the $1P$ band. The multiplicity is determined by the two $L=1$ orbital modes ($\rho$ and $\lambda$) and the spin channels allowed by permutation symmetry. For example, in the $\Lambda_c$ sector, both the $I=0$ light flavor wave function and the color wave function are antisymmetric, so Fermi statistics requires an antisymmetric spin-space wave function. The symmetric $\lambda_{Q-nn}$ mode therefore has $s_{nn}=0$, whereas the antisymmetric $\rho_{nn}$ mode has $s_{nn}=1$. The former allows only $S=\tfrac12$, while the latter allows $S=\tfrac12$ and $\tfrac32$. Therefore, the permutation symmetry gives $(3,3,1)$ states for $J^P=(\tfrac12^-,\tfrac32^-,\tfrac52^-)$, respectively. In the $\Xi_c(nsc)$ sector, all three quarks are distinct, so each orbital mode can combine with two independent $S=\tfrac12$ spin channels and one $S=\tfrac32$ channel, giving $(6,6,2)$ states for $J^P=(\tfrac12^-,\tfrac32^-,\tfrac52^-)$.

\begin{figure*}[htbp]
  \centering
  \includegraphics[width=1.0\textwidth]{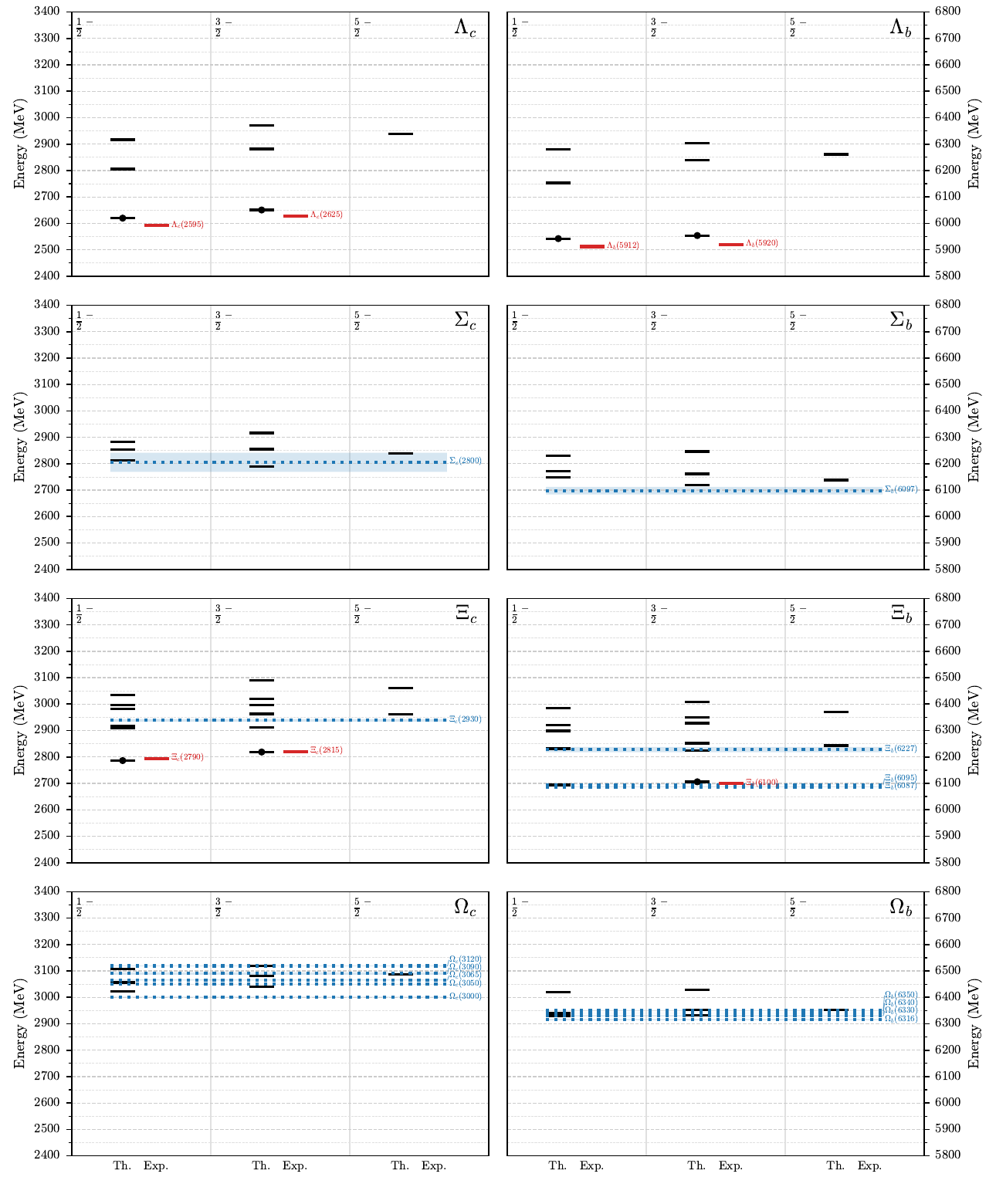}
  \caption{Computed $1P$ states (black) of the single-charm (left column) and single-bottom (right column) baryons compared with the experimental states. Within each panel the columns show the theoretical and experimental results for $J^P=\tfrac12^-,\tfrac32^-,\tfrac52^-$. The dots on the theoretical states denote the states included in the fit. Experimental results (PDG~\cite{ParticleDataGroup:2024cfk}) are shown as bands of full width $\Gamma$. Red solid lines denote states with assigned or measured $J^P$, and blue dotted lines denote states with undetermined $J^P$ (drawn spanning the full panel width).}
  \label{fig:cmp}
\end{figure*}

Figure~\ref{fig:cmp} compares the computed single-charm and single-bottom $1P$ states with the experimental results. The theoretical results agree well with the experimental results. The $\Xi_c(2790)$ and $\Xi_c(2815)$ are matched almost exactly, and the theoretical $\tfrac12^-$--$\tfrac32^-$ mass splitting reproduces the measured $\Xi_c(2790)$--$\Xi_c(2815)$ separation. The bottom counterpart $\Xi_b(6100)$ ($\tfrac32^-$) lies within $6$ MeV of the measured mass. The $\Lambda_c(2595)/\Lambda_c(2625)$ and $\Lambda_b(5912)/\Lambda_b(5920)$ doublets have the correct ordering but are $20$--$35$~MeV above the experimental values. These deviations are the largest in the excited heavy baryon sector. 
The $\Lambda_c(2595)$ lies almost exactly at the $\Sigma_c\pi$ threshold and is expected to receive a downward coupled-channel shift that the present model does not include. Other approaches show a similar pattern. The pre-observation predictions of Ref.~\cite{Karliner:2008sv} also place the $\Lambda_b(5912)/\Lambda_b(5920)$ doublet $15$--$20$~MeV above the measured masses. Further quark-model spectra for these states are given in Refs.~\cite{Ebert:2007nw,Ebert:2011kk,Roberts:2007ni,Valcarce:2008dr}.  The $\Sigma_{c,b}$, $\Xi_{c,b}$,  and $\Omega_{c,b}$ multiplets, whose spin-parity assignments have not yet been determined, are shown alongside the calculated states in Fig.~\ref{fig:cmp} for comparison.

The doubly- and triply-heavy $1P$ spectra are listed in Tables~\ref{tab:1P_QQq}--\ref{tab:1P_Q1Q1Q2}. The $\tfrac12^-$--$\tfrac32^-$ splitting decreases as the quark gets heavier---$31$~MeV for the $\Lambda_c$ doublet, $12$~MeV for the $\Lambda_b$ doublet, and below $2$~MeV for $\Xi_{cc}$ and $\Xi_{bb}$, whose lowest two states are nearly degenerate. 

The $2S$ radial spectrum is given in Table~\ref{tab:2S}. The radial gap $2S-1S$ decreases with increasing reduced mass, from $\sim590$~MeV in the single-charm sector to $\sim340$--$430$~MeV for the $bb$ and triply-heavy systems.

\subsubsection{Compositions and sizes}\label{sec:comp}
In addition to the mass spectrum, the wave function can characterize the internal structure of the calculated states. We write the normalized solution of Eq.~\eqref{eq:psi} as
\begin{equation}
  |\Psi\rangle=\sum_\alpha c_\alpha|\phi_\alpha\rangle,
\end{equation}
where $\alpha$ labels the (antisymmetrized) basis and 
\begin{equation}
  N_{\alpha\beta}=\langle\phi_\alpha|\phi_\beta\rangle
\end{equation}
is the overlap matrix, with the normalization
\begin{equation}
  \sum_{\alpha\beta}c_\alpha^{*}N_{\alpha\beta}c_\beta=1.
\end{equation} 
The probability that the wave function belongs to a class $X$ is then
\begin{equation}\label{eq:pop}
  P(X)=\sum_{\alpha\in X}\sum_{\beta}c_\alpha^{*}\,N_{\alpha\beta}\,c_\beta.
\end{equation}
The probabilities $P(X)$ sum to unity when $X$ runs over a complete partition of the basis. For the spin probabilities $P(S)$, $X$ collects the channels of total quark spin $S=\tfrac12$ or $\tfrac32$. Basis functions of different $S$ are orthogonal, so 
\begin{equation}
  P(S)=\langle\Psi|\hat{\mathcal P}_{S}|\Psi\rangle
\end{equation}
is an exact projection probability. However, for the orbital probabilities, $X$ collects the components in which the quark pair $(ab)$ carries the orbital angular momentum [$P(l_{ab}{=}1)$, a $\rho$ mode] or the spectator $c$ orbits that pair [$P(l_{c\text{-}ab}{=}1)$, a $\lambda$ mode]. Basis functions belonging to different Jacobi channels in Fig.~\ref{fig:jacobi} are not orthogonal. Therefore, the orbital probabilities sum to unity by Eq.~\eqref{eq:pop}, but the individual values may fall outside $[0,1]$ when the channels overlap strongly. They should be interpreted as qualitative indicators. This prescription is mathematically equivalent to the Chirgwin--Coulson weights and the Mulliken partition commonly used in quantum chemistry~\cite{Chirgwin_1950,Mulliken_1955}.

As a quantitative indicator of the spatial structure, we also evaluate the pairwise root-mean-square radii
\begin{equation}\label{eq:rms}
  r_{ab}\equiv\langle\Psi|\,\bm{r}_{ab}^{\,2}\,|\Psi\rangle^{1/2},
\end{equation}
where $\bm{r}_{ab}=\bm{r}_a-\bm{r}_b$. 

The probabilities and radii of excited heavy baryons are listed in Tables~\ref{tab:1P_Qnn}--\ref{tab:1P_Q1Q1Q2}. The sizes of the excited baryons range from $0.3$ to $1.2$~fm. These baryons are sufficiently compact for the nonrelativistic treatment to be self-consistent.
 
As shown in Tables~\ref{tab:1P_Qnn}--\ref{tab:1P_Q1Q1Q2}, every $\tfrac52^-$ state has $P(S=\tfrac32)=1$, since $S=\tfrac12$ cannot couple with $L=1$ to $J=\tfrac52$. The lowest $\Lambda_c(\tfrac12^-,\tfrac32^-)$ and $\Lambda_b(\tfrac12^-,\tfrac32^-)$ doublets have $P(S=\tfrac12)=1.00$. Their $\Sigma_Q$ counterparts are strongly mixed instead. For the lowest two $\Sigma_c(\tfrac12^-)$ states, the values of $(P(S=\tfrac12),P(S=\tfrac32))$ are $(0.61,0.39)$ and $(0.39,0.61)$, respectively.

The pairwise radii and the orbital probabilities give a consistent picture of the $\rho$ and $\lambda$ modes. 
In the singly-heavy sector, the lowest states are predominantly $\lambda$ mode excitations between the light and heavy degrees of freedom. This pattern is expected in the heavy-quark limit. The $P$-wave centrifugal potential decreases as the reduced mass increases. The larger reduced mass of the $\lambda_{Q-nn}$ coordinate therefore reduces the gap between the $P$- and $S$-wave states. The lowest $\Lambda_c(\tfrac12^-)$ state has a compact light pair ($r_{nn}=0.73$~fm) with the charm quark orbiting further out ($r_{nc}=0.85$~fm). The two radii exchange roles in the $\rho$ mode state at $2806$~MeV ($r_{nn}=0.96$~fm, $r_{nc}=0.74$~fm). The $\Lambda_b$ doublet repeats the pattern. Replacing the light pair by $ss$ makes the system nearly equilateral. The $\Omega_c$ states have $r_{ss}\simeq r_{sc}\simeq0.73$--$0.76$~fm, and the lowest $\Xi_c(nsc,\tfrac12^-)$ state, in which all three quarks are distinct, likewise has $r_{ns}=0.71$, $r_{nc}=0.79$, $r_{sc}=0.77$~fm.

The doubly-heavy systems invert the ordering, because the heavy pair, with its larger reduced mass, is the softer degree of freedom. The qualitative orbital probabilities also support a $\rho_{QQ}$ mode assignment for the lowest doubly-heavy states, and the pairwise radii provide a direct geometric picture. The lowest $\Xi_{cc}$ $1P$ states have a stretched heavy pair, $r_{cc}=0.66$~fm, whereas the states near $4.1$~GeV have a compact $cc$ diquark with the light quark far outside ($r_{cc}=0.50$~fm, $r_{cn}=0.92$~fm). The same separation of scales is clearer in $\Xi_{bb}$, where the diquark contracts from $r_{bb}=0.46$~fm in the lowest doublet to $0.31$~fm in the upper states while the light quark moves out to $0.90$~fm. 

\subsection{Excited light baryons}\label{sec:exc_light}

\begin{figure*}[tbp]
  \centering
  \includegraphics[width=0.98\textwidth]{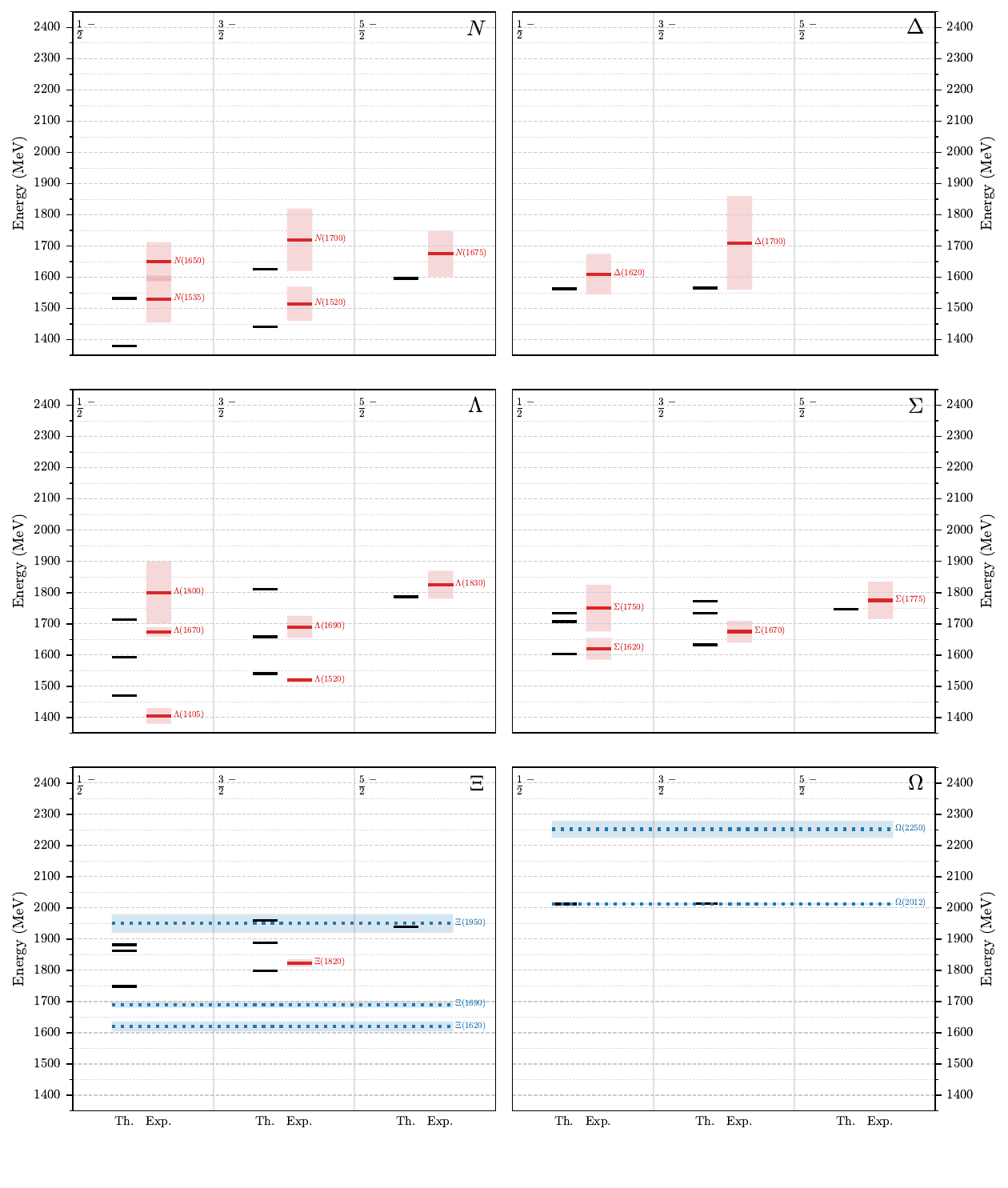}
  \caption{Computed $1P$ states (black) of the light baryons compared with the experimental states. Within each panel the columns show the theoretical and experimental results for $J^P=\tfrac12^-,\tfrac32^-,\tfrac52^-$. Experimental results (PDG~\cite{ParticleDataGroup:2024cfk}) are shown as bands of full width $\Gamma$. Red solid lines denote states with assigned or measured $J^P$, and blue dotted lines denote states with undetermined $J^P$ (drawn spanning the full panel width). These states are not included in the fit.}
  \label{fig:cmp_light}
\end{figure*}

We apply the same approach to the excited light baryons. Although these states are not part of the fit for the reasons given in Sec.~\ref{sec:fit}, they are calculated with the same parameters. It is worth studying how well the constituent quark model works in this sector. The $1P$ states with their compositions and sizes are listed in Tables~\ref{tab:1P_Lqqq} and~\ref{tab:1P_Lqqp}, and the $2S$ states are listed in Table~\ref{tab:2S_light}. Figure~\ref{fig:cmp_light} compares the $1P$ spectra with the experimental results.

\subsubsection{Spectrum}\label{sec:exc_light_spec}

The agreement between theoretical results and experimental data in the light sector is notably poorer than in the heavy sector. The deviations follow a clear pattern. The nonstrange negative-parity states are systematically low. The two nucleon $\tfrac12^-$ states lie at $1380$ and $1532$~MeV against the $N(1535)$ and $N(1650)$, the $\tfrac32^-$ pair lies at $1441$ and $1626$~MeV against the $N(1520)$ and $N(1700)$, and the $\tfrac52^-$ state lies at $1596$~MeV against the $N(1675)$, with deviations from $-74$ to $-150$~MeV. The two $\Delta$ states behave similarly, with $-47$ and $-144$~MeV deviations. The radial excitations deviate in the opposite direction, lying $200$--$340$~MeV above the corresponding candidates, as shown in Table~\ref{tab:2S_light}. In the nucleon spectrum, the present quark potential model reverses the ordering of the $2S$ and $1P$ states, whereas the $N(1440)$ is observed below the $N(1520)$. The lowest calculated $\Lambda(\tfrac12^-)$ is $1470$~MeV, which is $65$~MeV above the experimental $\Lambda(1405)$. The lowest calculated $\Xi$ is $1748$~MeV. It is $128$ and $58$~MeV above the $\Xi(1620)$ and $\Xi(1690)$, respectively.

We believe that it is difficult to repair these deviations by readjusting the parameter values. It is more natural to associate them with physics that the conventional quark model does not contain. The $N(1440)$, $N(1535)$, $\Lambda(1405)$, $\Xi(1620)$ and $\Xi(1690)$ are the standard examples in which coupled-channel and meson-cloud dynamics are believed to play a pivotal role~\cite{Chen:2016spr,Guo:2017jvc,Meng:2022ozq}. In contrast, the $\Lambda(1520)$, $\Sigma(1670)$, $\Sigma(1775)$, $\Lambda(1830)$, and $\Xi(1820)$ are reproduced to within about $40$~MeV, the same level of accuracy that the model achieves for the heavy baryons. 

The excited light sector thus empirically delimits the region in which we would trust the present quark potential model. It performs well for heavy hadrons, but works poorly where light degrees of freedom and the coupled-channel effects become important. This is also why we prefer to leave these states out of the fitting process. Including them would let the fit absorb dynamics that the model cannot represent, at the expense of its performance in the sectors it describes well.

\subsubsection{Compositions and sizes}\label{sec:comp_light}
The internal structure, obtained from Eqs.~(\ref{eq:pop}) and~(\ref{eq:rms}) in Sec.~\ref{sec:comp}, is less clearly organized than in the heavy sector. With three quarks of comparable mass, there is no heavy-quark limit to separate the $\rho$ and $\lambda$ excitations. For the nucleon, the orbital probabilities range from $\rho$-dominated ($0.67$ at $1380$~MeV) to $\lambda$-dominated ($0.94$ at $1532$~MeV) within the same $J^P$. The two $\Delta$ states each show an even admixture ($0.56$ and $0.55$). The spin content is likewise mixed. The lowest $N(\tfrac12^-)$ has $P(S=\tfrac12)=0.78$, while for the $\Delta$ and the $\Omega$, whose flavor wave functions are symmetric, the $1P$ states are pure $S=\tfrac12$.

The radii show the expected scale hierarchy. The light $1P$ baryons are $0.8$--$1.2$~fm objects, roughly $50\%$ larger than their heavy counterparts. The excited nucleon states have radii $r_{nn}=0.93$--$1.12$~fm. These radii decrease monotonically with increasing strangeness. The $\Omega$ $1P$ states are the most compact in the light sector with $r_{ss}=0.86$~fm, approaching the size of the singly-charmed baryons. In the $\Lambda$ system, the $nn$ separation is smaller than the $ns$ separation ($r_{nn}=0.76$~fm versus $r_{ns}=0.90$~fm in the lowest $\tfrac12^-$ state). The same diquark-like arrangement is seen in the $\Lambda_c$ and $\Lambda_b$. This feature of the heavy $\Lambda_Q$ systems survives into the light sector even though the strange quark mass is not as heavy as the charm and bottom quarks.

\section{Summary}\label{sec:summary}
The original AL1 quark potential uses a single parameter set for mesons and baryons of all flavors. We have augmented it with the spin--orbit and tensor interactions. We refit the model parameters to the latest experimental data. The data set comprises the well-established hadrons whose interpretation as simple $q\bar q$ or $qqq$ states is uncontroversial. We deliberately exclude the excited light hadrons and some typical exotic hadron candidates such as $N(1440)$, $N(1535)$, $\Lambda(1405)$, $a_0(980)$, $f_0(980)$, $D_{s0}^*(2317)$, and $D_{s1}(2460)$ from the global fit.

The fit reaches $\chi^2/\mathrm{dof}=1.04$, with the vast majority of deviations below $20$ MeV. Using the refitted parameters, we predict the unobserved doubly- and triply-heavy $1S$ baryons. We further determine the spectra, compositions, and sizes of excited heavy baryons. Finally, we calculate the excited light baryon spectrum, which is not included in the fit. 

The newly observed $\Omega_{cc}$ is not included in our data set. Its measured mass lies within about $1$~MeV of the predicted value. 

For the excited heavy baryon spectra, the theoretical predictions agree well with the experimental results. The exact spin projections give $P(S=\tfrac12)=1.00$ for the lowest $\Lambda_Q$ doublets, whereas the $\Sigma_Q$ states are strongly mixed. The qualitative orbital probabilities support a $\lambda$ mode assignment for the lowest singly-heavy states and a $\rho$ mode assignment for the lowest doubly-heavy states. The pairwise radii provide a direct geometric picture. 

We apply the refitted Hamiltonian to light baryons without fitting them. For the nucleon, it places the negative-parity states too low and the radial excitations too high. The calculated $N(1440)$ and $\Lambda(1405)$ masses still deviate from experiment by $235$ and $65$~MeV, respectively. Our present model does not resolve this long-standing problem. This pattern marks the boundary at which coupled-channel dynamics takes over from the constituent picture.

The present model treats every flavor sector with a single parameter set. It covers both mesons and baryons, as well as orbital and radial excitations. It therefore provides a controlled starting point for few-body calculations of multiquark states. These applications require a reliable description of all constituent two- and three-body subsystems. We urge experimental searches for the predicted states. Such measurements would test the model and deepen our understanding of the strong interaction in the hadron spectrum.

\begin{acknowledgments}
This project was supported by the National Natural Science Foundation of China (Grant No.~12475137) and Start-up Funds of Southeast University (Grant No. 4007022506). The computational resources were supported by the High-performance Computing Platform of Peking University. 
\end{acknowledgments}

\section*{Data Availability}
The data that support the findings of this article are available within the article.

\appendix
\section{Excited baryon spectra, compositions, and sizes}\label{app:spectra}
For completeness, we collect here the full excited baryon results discussed in Sec.~\ref{sec:result}: the $1P$ eigenvalues with their spin--orbital compositions and pairwise sizes [Eqs.~(\ref{eq:pop}) and~(\ref{eq:rms})], organized by flavor-symmetry class (Tables~\ref{tab:1P_Qnn}--\ref{tab:1P_Q1Q1Q2} for the heavy sector, Tables~\ref{tab:1P_Lqqq} and~\ref{tab:1P_Lqqp} for the light sector), and the $2S$ radial spectra (Tables~\ref{tab:2S} and~\ref{tab:2S_light}). The notations are defined in Table~\ref{tab:1P_Qnn} and used throughout. The light baryon entries are not part of the fit; they are included for the discussion in Sec.~\ref{sec:exc_light}.

\begin{table*}[p]
\centering\small\setlength{\tabcolsep}{4pt}
\caption{$1P$ ($L=1$) eigenstates and compositions for the $Qnn$ (one heavy $Q$ + two $u/d$) baryons. The lowest $(3,3,1)$ states for $J^P=(\tfrac12^-,\tfrac32^-,\tfrac52^-)$ are listed. The multiplicity is fixed by permutation symmetry. The higher levels belong to the $2P$ states. Columns: energy (MeV); the spin probabilities $P(S{=}\tfrac12)$, $P(S{=}\tfrac32)$ and the orbital probabilities, both given by Eq.~\eqref{eq:pop} and labeled by quark content --- $P(l_{ab}{=}1)$ is the probability that the $(ab)$ pair carries the unit orbital angular momentum (a $\rho$ excitation), and $P(l_{c\text{-}ab}{=}1)$ that the spectator orbits the $(ab)$ pair (a $\lambda$ excitation); the pairwise root-mean-square separations $r_{ab}$ [Eq.~\eqref{eq:rms}] in fm. An asterisk marks the levels included in the fit. Quark labels: $Q=c$ ($\Lambda_c,\Sigma_c$), $b$ ($\Lambda_b,\Sigma_b$). The orbital probabilities should be interpreted as a qualitative indicator.}
\label{tab:1P_Qnn}
\begin{tabular}{ll c cc cccccc}
\toprule
Baryon & $I(J^P)$ & $E$~[MeV] & $P(S{=}\tfrac12)$ & $P(S{=}\tfrac32)$ & $P(l_{nn})$ & $P(l_{nQ})$ & $P(l_{Q\text{-}nn})$ & $P(l_{n\text{-}nQ})$ & $r_{nn}$ & $r_{nQ}$ \\
\midrule
$\Lambda_c$ & $0(\tfrac12^-)$ & 2620$^{*}$ & 1.00 & 0.00 & $-0.00$ & $+0.05$ & $+0.52$ & $+0.43$ & $0.73$ & $0.85$ \\
 &  & 2806 & 0.46 & 0.54 & $+0.43$ & $-0.06$ & $+0.00$ & $+0.63$ & $0.96$ & $0.74$ \\
 &  & 2917 & 0.55 & 0.45 & $+0.32$ & $-0.50$ & $+0.02$ & $+1.17$ & $1.11$ & $0.81$ \\
 & $0(\tfrac32^-)$ & 2651$^{*}$ & 1.00 & 0.00 & $+0.00$ & $-0.16$ & $+0.57$ & $+0.59$ & $0.75$ & $0.87$ \\
 &  & 2882 & 0.94 & 0.06 & $+0.12$ & $-0.36$ & $+0.01$ & $+1.24$ & $1.18$ & $0.85$ \\
 &  & 2972 & 0.06 & 0.94 & $+0.25$ & $-0.08$ & $-0.00$ & $+0.84$ & $1.15$ & $0.84$ \\
 & $0(\tfrac52^-)$ & 2939 & 0.00 & 1.00 & $+0.21$ & $-0.48$ & $+0.00$ & $+1.26$ & $1.22$ & $0.88$ \\
\midrule
$\Sigma_c$ & $1(\tfrac12^-)$ & 2813 & 0.61 & 0.39 & $-0.01$ & $+1.22$ & $-0.14$ & $-0.06$ & $0.88$ & $0.87$ \\
 &  & 2854 & 0.39 & 0.61 & $+0.01$ & $-2.29$ & $+0.80$ & $+2.49$ & $0.90$ & $0.89$ \\
 &  & 2883 & 1.00 & 0.00 & $+0.27$ & $-0.18$ & $+0.03$ & $+0.88$ & $1.11$ & $0.83$ \\
 & $1(\tfrac32^-)$ & 2790 & 0.95 & 0.05 & $+0.01$ & $-0.30$ & $+0.30$ & $+1.00$ & $0.92$ & $0.90$ \\
 &  & 2855 & 0.18 & 0.82 & $+0.05$ & $+0.43$ & $+0.19$ & $+0.33$ & $0.95$ & $0.91$ \\
 &  & 2916 & 0.87 & 0.13 & $+0.21$ & $+0.06$ & $-0.00$ & $+0.74$ & $1.09$ & $0.84$ \\
 & $1(\tfrac52^-)$ & 2840 & 0.00 & 1.00 & $+0.00$ & $-0.29$ & $+0.38$ & $+0.91$ & $0.92$ & $0.94$ \\
\midrule
$\Lambda_b$ & $0(\tfrac12^-)$ & 5942$^{*}$ & 1.00 & 0.00 & $+0.00$ & $-0.06$ & $+0.35$ & $+0.72$ & $0.74$ & $0.81$ \\
 &  & 6153 & 0.37 & 0.63 & $+0.26$ & $-0.05$ & $+0.00$ & $+0.79$ & $0.96$ & $0.72$ \\
 &  & 6281 & 0.64 & 0.36 & $-0.04$ & $-0.39$ & $+0.01$ & $+1.42$ & $1.12$ & $0.80$ \\
 & $0(\tfrac32^-)$ & 5954$^{*}$ & 1.00 & 0.00 & $+0.00$ & $-0.11$ & $+0.35$ & $+0.76$ & $0.74$ & $0.82$ \\
 &  & 6239 & 0.89 & 0.11 & $-0.26$ & $+0.00$ & $+0.00$ & $+1.26$ & $1.19$ & $0.83$ \\
 &  & 6303 & 0.11 & 0.89 & $-0.06$ & $-0.42$ & $-0.00$ & $+1.48$ & $1.13$ & $0.81$ \\
 & $0(\tfrac52^-)$ & 6261 & 0.00 & 1.00 & $-0.25$ & $+0.02$ & $+0.00$ & $+1.23$ & $1.20$ & $0.85$ \\
\midrule
$\Sigma_b$ & $1(\tfrac12^-)$ & 6148 & 0.65 & 0.35 & $+0.00$ & $-1.32$ & $-0.86$ & $+3.18$ & $0.90$ & $0.86$ \\
 &  & 6173 & 0.36 & 0.64 & $-0.00$ & $+2.86$ & $+1.47$ & $-3.33$ & $0.89$ & $0.85$ \\
 &  & 6230 & 0.99 & 0.01 & $-0.08$ & $+0.06$ & $-0.02$ & $+1.03$ & $1.11$ & $0.79$ \\
 & $1(\tfrac32^-)$ & 6119 & 0.91 & 0.09 & $-0.00$ & $-0.02$ & $+0.01$ & $+1.02$ & $0.91$ & $0.87$ \\
 &  & 6162 & 0.16 & 0.84 & $-0.00$ & $-1.30$ & $-0.72$ & $+3.02$ & $0.92$ & $0.87$ \\
 &  & 6246 & 0.93 & 0.07 & $-0.07$ & $+0.00$ & $-0.09$ & $+1.16$ & $1.10$ & $0.80$ \\
 & $1(\tfrac52^-)$ & 6138 & 0.00 & 1.00 & $+0.00$ & $-0.07$ & $+0.02$ & $+1.06$ & $0.91$ & $0.88$ \\
\bottomrule
\end{tabular}
\end{table*}

\begin{table*}[p]
\centering\small\setlength{\tabcolsep}{4pt}
\caption{$1P$ ($L=1$) eigenstates and compositions for the $Qns$ (one heavy $Q$ $+$ $u/d$ $+$ $s$) baryons [$Q=c$ ($\Xi_c$), $b$ ($\Xi_b$)]. For each $J^P$, the lowest $(6,6,2)$ states are listed. An asterisk marks the levels included in the fit. Columns and conventions as in Table~\ref{tab:1P_Qnn}.}
\label{tab:1P_Qns}
\begin{tabular}{ll c cc ccccccccc}
\toprule
Baryon & $I(J^P)$ & $E$~[MeV] & $P(S{=}\tfrac12)$ & $P(S{=}\tfrac32)$ & $P(l_{ns})$ & $P(l_{nQ})$ & $P(l_{sQ})$ & $P(l_{Q\text{-}ns})$ & $P(l_{s\text{-}nQ})$ & $P(l_{n\text{-}sQ})$ & $r_{ns}$ & $r_{nQ}$ & $r_{sQ}$ \\
\midrule
$\Xi_c$ & $\tfrac12(\tfrac12^-)$ & 2786$^{*}$ & 0.99 & 0.01 & $+0.01$ & $+0.06$ & $+0.70$ & $+0.68$ & $-0.43$ & $-0.02$ & $0.71$ & $0.79$ & $0.77$ \\
 &  & 2908 & 0.92 & 0.08 & $+0.10$ & $-0.37$ & $+0.90$ & $+0.44$ & $-0.41$ & $+0.34$ & $0.85$ & $0.73$ & $0.80$ \\
 &  & 2918 & 0.11 & 0.89 & $-0.06$ & $-0.18$ & $-0.61$ & $+0.28$ & $+1.21$ & $+0.37$ & $0.84$ & $0.74$ & $0.80$ \\
 &  & 2982 & 0.47 & 0.53 & $+0.62$ & $+1.83$ & $-0.45$ & $-0.07$ & $+0.06$ & $-0.98$ & $0.86$ & $0.92$ & $0.63$ \\
 &  & 2997 & 0.86 & 0.14 & $+0.13$ & $-0.17$ & $+0.05$ & $+0.13$ & $+0.28$ & $+0.59$ & $1.02$ & $0.89$ & $0.67$ \\
 &  & 3035 & 0.65 & 0.35 & $+0.44$ & $-0.28$ & $+0.13$ & $+0.01$ & $-0.09$ & $+0.80$ & $1.02$ & $0.83$ & $0.66$ \\
 & $\tfrac12(\tfrac32^-)$ & 2819$^{*}$ & 1.00 & 0.00 & $+0.01$ & $+0.06$ & $+0.71$ & $+0.72$ & $-0.47$ & $-0.03$ & $0.72$ & $0.81$ & $0.80$ \\
 &  & 2913 & 0.97 & 0.03 & $+0.01$ & $+0.11$ & $+0.58$ & $+0.38$ & $-0.10$ & $+0.01$ & $0.84$ & $0.82$ & $0.82$ \\
 &  & 2963 & 0.05 & 0.95 & $+0.08$ & $-0.60$ & $+2.00$ & $+0.69$ & $-1.52$ & $+0.35$ & $0.84$ & $0.83$ & $0.84$ \\
 &  & 2997 & 0.99 & 0.01 & $+0.34$ & $+0.07$ & $+0.23$ & $+0.00$ & $-0.07$ & $+0.43$ & $1.07$ & $0.88$ & $0.67$ \\
 &  & 3020 & 0.74 & 0.26 & $+0.45$ & $+0.06$ & $+0.19$ & $+0.03$ & $-0.15$ & $+0.41$ & $1.06$ & $0.90$ & $0.68$ \\
 &  & 3091 & 0.25 & 0.75 & $+0.26$ & $-0.19$ & $-0.02$ & $+0.06$ & $+0.20$ & $+0.70$ & $1.03$ & $0.90$ & $0.66$ \\
 & $\tfrac12(\tfrac52^-)$ & 2961 & 0.00 & 1.00 & $+0.00$ & $+0.17$ & $+0.58$ & $+0.45$ & $-0.16$ & $-0.05$ & $0.84$ & $0.86$ & $0.86$ \\
 &  & 3060 & 0.00 & 1.00 & $+0.36$ & $+0.11$ & $+0.20$ & $+0.01$ & $-0.10$ & $+0.42$ & $1.11$ & $0.93$ & $0.70$ \\
\midrule
$\Xi_b$ & $\tfrac12(\tfrac12^-)$ & 6094 & 1.00 & 0.00 & $+0.00$ & $+0.32$ & $+1.93$ & $+0.69$ & $-1.63$ & $-0.30$ & $0.70$ & $0.76$ & $0.73$ \\
 &  & 6230 & 0.63 & 0.37 & $-0.27$ & $+0.22$ & $-3.80$ & $+0.28$ & $+4.16$ & $+0.42$ & $0.84$ & $0.68$ & $0.78$ \\
 &  & 6233 & 0.37 & 0.63 & $+0.05$ & $-1.58$ & $+3.31$ & $+0.57$ & $-2.80$ & $+1.45$ & $0.81$ & $0.76$ & $0.77$ \\
 &  & 6299 & 0.39 & 0.61 & $+0.83$ & $+6.05$ & $+0.16$ & $+0.12$ & $-0.43$ & $-5.73$ & $0.88$ & $0.86$ & $0.55$ \\
 &  & 6321 & 0.90 & 0.10 & $+0.38$ & $+0.32$ & $+0.07$ & $+0.02$ & $+0.18$ & $+0.03$ & $1.03$ & $0.84$ & $0.63$ \\
 &  & 6385 & 0.72 & 0.28 & $+0.29$ & $-0.36$ & $+0.35$ & $+0.02$ & $-0.24$ & $+0.93$ & $1.02$ & $0.83$ & $0.62$ \\
 & $\tfrac12(\tfrac32^-)$ & 6106$^{*}$ & 1.00 & 0.00 & $+0.00$ & $+0.36$ & $+2.04$ & $+0.71$ & $-1.76$ & $-0.35$ & $0.71$ & $0.77$ & $0.74$ \\
 &  & 6223 & 0.94 & 0.06 & $-0.00$ & $+0.70$ & $+1.28$ & $+0.49$ & $-0.88$ & $-0.57$ & $0.82$ & $0.79$ & $0.77$ \\
 &  & 6252 & 0.07 & 0.93 & $+0.08$ & $-2.03$ & $+4.09$ & $+0.60$ & $-3.54$ & $+1.80$ & $0.82$ & $0.79$ & $0.78$ \\
 &  & 6328 & 0.88 & 0.12 & $+0.44$ & $+0.77$ & $+0.57$ & $+0.00$ & $-0.37$ & $-0.42$ & $1.04$ & $0.84$ & $0.64$ \\
 &  & 6349 & 0.81 & 0.19 & $+0.38$ & $+1.28$ & $+0.61$ & $+0.01$ & $-0.43$ & $-0.86$ & $1.08$ & $0.87$ & $0.65$ \\
 &  & 6409 & 0.31 & 0.69 & $+0.28$ & $-0.27$ & $+0.25$ & $+0.03$ & $-0.12$ & $+0.82$ & $1.02$ & $0.85$ & $0.62$ \\
 & $\tfrac12(\tfrac52^-)$ & 6243 & 0.00 & 1.00 & $-0.01$ & $+0.82$ & $+1.32$ & $+0.52$ & $-0.96$ & $-0.70$ & $0.83$ & $0.81$ & $0.78$ \\
 &  & 6371 & 0.00 & 1.00 & $+0.39$ & $+1.31$ & $+0.66$ & $+0.01$ & $-0.47$ & $-0.90$ & $1.10$ & $0.88$ & $0.66$ \\
\bottomrule
\end{tabular}
\end{table*}

\begin{table*}[p]
\centering\small\setlength{\tabcolsep}{4pt}
\caption{$1P$ ($L=1$) eigenstates and compositions for the $Qss$ (one heavy $Q$ $+$ two $s$) baryons [$Q=c$ ($\Omega_c$), $b$ ($\Omega_b$)]. For each $J^P$, the lowest $(3,3,1)$ states are listed. Columns and conventions as in Table~\ref{tab:1P_Qnn}.}
\label{tab:1P_Qss}
\begin{tabular}{ll c cc cccccc}
\toprule
Baryon & $I(J^P)$ & $E$~[MeV] & $P(S{=}\tfrac12)$ & $P(S{=}\tfrac32)$ & $P(l_{ss})$ & $P(l_{sQ})$ & $P(l_{Q\text{-}ss})$ & $P(l_{s\text{-}sQ})$ & $r_{ss}$ & $r_{sQ}$ \\
\midrule
$\Omega_c$ & $0(\tfrac12^-)$ & 3022 & 0.58 & 0.42 & $+0.00$ & $-0.03$ & $+0.04$ & $+0.99$ & $0.73$ & $0.73$ \\
 &  & 3056 & 0.43 & 0.57 & $+0.01$ & $+1.28$ & $+0.86$ & $-1.15$ & $0.73$ & $0.76$ \\
 &  & 3107 & 0.98 & 0.02 & $+0.35$ & $+0.14$ & $-0.00$ & $+0.51$ & $0.91$ & $0.71$ \\
 & $0(\tfrac32^-)$ & 3040 & 0.96 & 0.04 & $+0.01$ & $+0.26$ & $+0.42$ & $+0.30$ & $0.75$ & $0.76$ \\
 &  & 3081 & 0.04 & 0.96 & $+0.00$ & $-0.30$ & $+0.33$ & $+0.98$ & $0.74$ & $0.78$ \\
 &  & 3119 & 1.00 & 0.00 & $+0.33$ & $+0.19$ & $+0.02$ & $+0.46$ & $0.93$ & $0.71$ \\
 & $0(\tfrac52^-)$ & 3087 & 0.00 & 1.00 & $+0.00$ & $+0.34$ & $+0.48$ & $+0.18$ & $0.75$ & $0.80$ \\
\midrule
$\Omega_b$ & $0(\tfrac12^-)$ & 6331 & 0.60 & 0.40 & $-0.00$ & $+4.53$ & $+0.00$ & $-3.52$ & $0.72$ & $0.70$ \\
 &  & 6341 & 0.40 & 0.60 & $+0.00$ & $-10.42$ & $+0.92$ & $+10.50$ & $0.72$ & $0.70$ \\
 &  & 6421 & 1.00 & 0.00 & $+0.25$ & $-0.58$ & $+0.00$ & $+1.32$ & $0.92$ & $0.66$ \\
 & $0(\tfrac32^-)$ & 6332 & 0.95 & 0.05 & $+0.00$ & $-0.94$ & $+0.39$ & $+1.54$ & $0.73$ & $0.72$ \\
 &  & 6353 & 0.05 & 0.95 & $+0.00$ & $+4.11$ & $+0.13$ & $-3.25$ & $0.73$ & $0.72$ \\
 &  & 6429 & 1.00 & 0.00 & $+0.25$ & $-0.69$ & $+0.00$ & $+1.44$ & $0.92$ & $0.67$ \\
 & $0(\tfrac52^-)$ & 6353 & 0.00 & 1.00 & $+0.00$ & $-1.08$ & $+0.43$ & $+1.65$ & $0.73$ & $0.73$ \\
\bottomrule
\end{tabular}
\end{table*}

\begin{table*}[p]
\centering\small\setlength{\tabcolsep}{4pt}
\caption{$1P$ ($L=1$) eigenstates and compositions for the $QQq$ (two identical heavy $+$ one light) baryons [$(Q,q)=(c,n)\,\Xi_{cc},\ (b,n)\,\Xi_{bb},\ (c,s)\,\Omega_{cc},\ (b,s)\,\Omega_{bb}$]. The lowest $(3,3,1)$ states are listed. Columns and conventions as in Table~\ref{tab:1P_Qnn}.}
\label{tab:1P_QQq}
\begin{tabular}{ll c cc cccccc}
\toprule
Baryon & $I(J^P)$ & $E$~[MeV] & $P(S{=}\tfrac12)$ & $P(S{=}\tfrac32)$ & $P(l_{QQ})$ & $P(l_{Qq})$ & $P(l_{q\text{-}QQ})$ & $P(l_{Q\text{-}Qq})$ & $r_{QQ}$ & $r_{Qq}$ \\
\midrule
$\Xi_{cc}$ & $\tfrac12(\tfrac12^-)$ & 3946 & 0.99 & 0.01 & $+0.35$ & $+0.05$ & $-0.01$ & $+0.62$ & $0.66$ & $0.72$ \\
 &  & 4072 & 0.93 & 0.07 & $+0.00$ & $+0.55$ & $+0.38$ & $+0.07$ & $0.50$ & $0.92$ \\
 &  & 4106 & 0.08 & 0.92 & $+0.02$ & $+0.50$ & $+0.39$ & $+0.08$ & $0.47$ & $0.86$ \\
 & $\tfrac12(\tfrac32^-)$ & 3948 & 1.00 & 0.00 & $+0.34$ & $+0.04$ & $+0.00$ & $+0.62$ & $0.67$ & $0.73$ \\
 &  & 4090 & 0.64 & 0.36 & $+0.00$ & $+0.40$ & $+0.48$ & $+0.12$ & $0.50$ & $0.94$ \\
 &  & 4172 & 0.36 & 0.64 & $-0.00$ & $+0.56$ & $+0.30$ & $+0.14$ & $0.49$ & $0.91$ \\
 & $\tfrac12(\tfrac52^-)$ & 4120 & 0.00 & 1.00 & $+0.00$ & $+0.42$ & $+0.51$ & $+0.07$ & $0.50$ & $0.97$ \\
\midrule
$\Xi_{bb}$ & $\tfrac12(\tfrac12^-)$ & 10428 & 1.00 & 0.00 & $+0.43$ & $+0.04$ & $+0.00$ & $+0.53$ & $0.46$ & $0.65$ \\
 &  & 10599 & 0.90 & 0.10 & $+0.00$ & $+0.35$ & $+0.51$ & $+0.14$ & $0.31$ & $0.90$ \\
 &  & 10659 & 0.10 & 0.90 & $+0.00$ & $-0.43$ & $+1.29$ & $+0.14$ & $0.30$ & $0.83$ \\
 & $\tfrac12(\tfrac32^-)$ & 10426 & 1.00 & 0.00 & $+0.50$ & $+0.00$ & $-0.00$ & $+0.50$ & $0.46$ & $0.66$ \\
 &  & 10607 & 0.58 & 0.42 & $+0.00$ & $+0.39$ & $+0.46$ & $+0.15$ & $0.31$ & $0.90$ \\
 &  & 10691 & 0.43 & 0.57 & $-0.00$ & $-0.25$ & $+1.08$ & $+0.18$ & $0.31$ & $0.85$ \\
 & $\tfrac12(\tfrac52^-)$ & 10620 & 0.00 & 1.00 & $+0.00$ & $+0.48$ & $+0.40$ & $+0.12$ & $0.31$ & $0.92$ \\
\midrule
$\Omega_{cc}$ & $0(\tfrac12^-)$ & 4059 & 0.95 & 0.05 & $+0.06$ & $+0.20$ & $-0.04$ & $+0.78$ & $0.63$ & $0.63$ \\
 &  & 4132 & 0.21 & 0.79 & $+0.00$ & $+0.58$ & $+0.33$ & $+0.08$ & $0.49$ & $0.73$ \\
 &  & 4159 & 0.84 & 0.16 & $+0.00$ & $+0.55$ & $+0.13$ & $+0.31$ & $0.50$ & $0.78$ \\
 & $0(\tfrac32^-)$ & 4071 & 1.00 & 0.00 & $+0.06$ & $+0.13$ & $+0.00$ & $+0.81$ & $0.66$ & $0.64$ \\
 &  & 4171 & 0.97 & 0.03 & $+0.00$ & $+0.69$ & $+0.15$ & $+0.16$ & $0.49$ & $0.79$ \\
 &  & 4187 & 0.03 & 0.97 & $+0.00$ & $+0.75$ & $+0.07$ & $+0.17$ & $0.49$ & $0.79$ \\
 & $0(\tfrac52^-)$ & 4203 & 0.00 & 1.00 & $+0.00$ & $+0.65$ & $+0.20$ & $+0.15$ & $0.50$ & $0.82$ \\
\midrule
$\Omega_{bb}$ & $0(\tfrac12^-)$ & 10524 & 1.00 & 0.00 & $-0.01$ & $+0.05$ & $+0.01$ & $+0.95$ & $0.45$ & $0.55$ \\
 &  & 10652 & 0.16 & 0.84 & $-0.00$ & $+0.69$ & $+0.35$ & $-0.03$ & $0.30$ & $0.69$ \\
 &  & 10670 & 0.84 & 0.16 & $+0.00$ & $+0.90$ & $+0.04$ & $+0.06$ & $0.31$ & $0.74$ \\
 & $0(\tfrac32^-)$ & 10526 & 1.00 & 0.00 & $+0.11$ & $+0.03$ & $-0.00$ & $+0.86$ & $0.45$ & $0.55$ \\
 &  & 10677 & 0.96 & 0.04 & $-0.00$ & $+0.85$ & $+0.10$ & $+0.05$ & $0.31$ & $0.74$ \\
 &  & 10682 & 0.04 & 0.96 & $-0.00$ & $+0.73$ & $+0.24$ & $+0.03$ & $0.30$ & $0.72$ \\
 & $0(\tfrac52^-)$ & 10692 & 0.00 & 1.00 & $+0.00$ & $+0.99$ & $-0.05$ & $+0.07$ & $0.31$ & $0.76$ \\
\bottomrule
\end{tabular}
\end{table*}

\begin{table*}[p]
\centering\small\setlength{\tabcolsep}{4pt}
\caption{$1P$ ($L=1$) eigenstates and compositions for the $Q_1Q_2q$ ($bc$ baryons, all three quarks distinct) [$q=n$ ($\Xi_{bc}$), $s$ ($\Omega_{bc}$)]; the lowest $(6,6,2)$ states are listed. Columns and conventions as in Table~\ref{tab:1P_Qnn}.}
\label{tab:1P_Q1Q2q}
\begin{tabular}{ll c cc ccccccccc}
\toprule
Baryon & $I(J^P)$ & $E$~[MeV] & $P(S{=}\tfrac12)$ & $P(S{=}\tfrac32)$ & $P(l_{bc})$ & $P(l_{cq})$ & $P(l_{bq})$ & $P(l_{q\text{-}bc})$ & $P(l_{c\text{-}bq})$ & $P(l_{b\text{-}cq})$ & $r_{bc}$ & $r_{cq}$ & $r_{bq}$ \\
\midrule
$\Xi_{bc}$ & $\tfrac12(\tfrac12^-)$ & 7156 & 0.99 & 0.01 & $+0.60$ & $-0.00$ & $-0.15$ & $+0.07$ & $-0.24$ & $+0.72$ & $0.57$ & $0.67$ & $0.68$ \\
 &  & 7213 & 0.99 & 0.01 & $+0.23$ & $+0.05$ & $-0.02$ & $+0.02$ & $+0.22$ & $+0.50$ & $0.58$ & $0.71$ & $0.68$ \\
 &  & 7217 & 0.02 & 0.98 & $+0.56$ & $+0.09$ & $+0.13$ & $-0.03$ & $-0.21$ & $+0.47$ & $0.58$ & $0.71$ & $0.69$ \\
 &  & 7358 & 0.91 & 0.09 & $-0.01$ & $+0.59$ & $-0.62$ & $+0.93$ & $+0.06$ & $+0.05$ & $0.42$ & $0.93$ & $0.89$ \\
 &  & 7394 & 0.33 & 0.67 & $-0.01$ & $+0.55$ & $-0.67$ & $+1.09$ & $-0.01$ & $+0.04$ & $0.39$ & $0.85$ & $0.84$ \\
 &  & 7426 & 0.76 & 0.24 & $+0.07$ & $+0.63$ & $-0.81$ & $+1.17$ & $-0.09$ & $+0.04$ & $0.41$ & $0.89$ & $0.84$ \\
 & $\tfrac12(\tfrac32^-)$ & 7169 & 0.99 & 0.01 & $+0.51$ & $+0.00$ & $-0.12$ & $+0.06$ & $-0.16$ & $+0.70$ & $0.58$ & $0.67$ & $0.68$ \\
 &  & 7215 & 0.99 & 0.01 & $+0.25$ & $-0.01$ & $-0.16$ & $+0.09$ & $+0.30$ & $+0.52$ & $0.59$ & $0.71$ & $0.69$ \\
 &  & 7232 & 0.02 & 0.98 & $-0.05$ & $-0.02$ & $-0.14$ & $+0.06$ & $+0.58$ & $+0.57$ & $0.59$ & $0.72$ & $0.70$ \\
 &  & 7358 & 0.92 & 0.08 & $+0.02$ & $+0.60$ & $-0.79$ & $+1.14$ & $-0.00$ & $+0.04$ & $0.41$ & $0.95$ & $0.90$ \\
 &  & 7374 & 0.68 & 0.32 & $-0.00$ & $+0.50$ & $-0.43$ & $+0.83$ & $+0.05$ & $+0.04$ & $0.42$ & $0.95$ & $0.90$ \\
 &  & 7454 & 0.40 & 0.60 & $+0.04$ & $+0.59$ & $-0.58$ & $+0.94$ & $-0.05$ & $+0.06$ & $0.42$ & $0.90$ & $0.86$ \\
 & $\tfrac12(\tfrac52^-)$ & 7237 & 0.00 & 1.00 & $+0.32$ & $-0.00$ & $-0.14$ & $+0.08$ & $+0.20$ & $+0.53$ & $0.60$ & $0.72$ & $0.71$ \\
 &  & 7392 & 0.00 & 1.00 & $-0.01$ & $+0.53$ & $-0.63$ & $+0.99$ & $+0.06$ & $+0.05$ & $0.43$ & $0.97$ & $0.92$ \\
\midrule
$\Omega_{bc}$ & $0(\tfrac12^-)$ & 7269 & 0.99 & 0.01 & $-0.08$ & $+0.01$ & $+0.10$ & $-0.01$ & $+0.50$ & $+0.49$ & $0.55$ & $0.57$ & $0.59$ \\
 &  & 7317 & 0.99 & 0.01 & $+0.19$ & $+0.02$ & $+0.09$ & $-0.05$ & $+0.33$ & $+0.42$ & $0.56$ & $0.61$ & $0.58$ \\
 &  & 7319 & 0.02 & 0.98 & $+0.57$ & $-0.01$ & $-0.01$ & $-0.03$ & $+0.05$ & $+0.43$ & $0.56$ & $0.61$ & $0.58$ \\
 &  & 7402 & 0.41 & 0.59 & $-0.02$ & $+0.47$ & $-0.02$ & $+0.45$ & $+0.06$ & $+0.05$ & $0.39$ & $0.72$ & $0.70$ \\
 &  & 7430 & 0.79 & 0.21 & $+0.02$ & $+0.36$ & $+0.15$ & $+0.37$ & $+0.09$ & $+0.00$ & $0.42$ & $0.77$ & $0.68$ \\
 &  & 7436 & 0.80 & 0.20 & $+0.04$ & $+0.42$ & $+0.24$ & $+0.23$ & $+0.02$ & $+0.05$ & $0.42$ & $0.79$ & $0.73$ \\
 & $0(\tfrac32^-)$ & 7283 & 0.99 & 0.01 & $-0.10$ & $+0.00$ & $+0.10$ & $-0.02$ & $+0.51$ & $+0.51$ & $0.56$ & $0.57$ & $0.59$ \\
 &  & 7324 & 1.00 & 0.00 & $+0.05$ & $+0.01$ & $+0.06$ & $+0.01$ & $+0.45$ & $+0.42$ & $0.57$ & $0.61$ & $0.60$ \\
 &  & 7338 & 0.01 & 0.99 & $-0.15$ & $+0.03$ & $+0.16$ & $-0.04$ & $+0.54$ & $+0.45$ & $0.57$ & $0.62$ & $0.60$ \\
 &  & 7437 & 0.95 & 0.05 & $+0.02$ & $+0.42$ & $+0.07$ & $+0.42$ & $+0.05$ & $+0.02$ & $0.42$ & $0.80$ & $0.74$ \\
 &  & 7450 & 1.00 & 0.00 & $+0.01$ & $+0.41$ & $+0.12$ & $+0.36$ & $+0.06$ & $+0.03$ & $0.42$ & $0.80$ & $0.73$ \\
 &  & 7457 & 0.05 & 0.95 & $+0.04$ & $+0.44$ & $+0.25$ & $+0.23$ & $+0.02$ & $+0.03$ & $0.42$ & $0.79$ & $0.72$ \\
 & $0(\tfrac52^-)$ & 7346 & 0.00 & 1.00 & $+0.09$ & $+0.00$ & $+0.06$ & $+0.01$ & $+0.40$ & $+0.44$ & $0.58$ & $0.62$ & $0.61$ \\
 &  & 7471 & 0.00 & 1.00 & $+0.01$ & $+0.43$ & $+0.01$ & $+0.47$ & $+0.06$ & $+0.02$ & $0.43$ & $0.83$ & $0.75$ \\
\bottomrule
\end{tabular}
\end{table*}

\begin{table*}[p]
\centering\small\setlength{\tabcolsep}{4pt}
\caption{$1P$ ($L=1$) eigenstates and compositions for the $QQQ$ (three identical heavy) baryons [$Q=c$ ($\Omega_{ccc}$), $b$ ($\Omega_{bbb}$)]. The $1P$ $\Omega_{QQQ}$ contains only one $\tfrac12^-$ and one $\tfrac32^-$ state (no $\tfrac52^-$). Columns and conventions as in Table~\ref{tab:1P_Qnn}.}
\label{tab:1P_Q3}
\begin{tabular}{ll c cc ccc}
\toprule
Baryon & $I(J^P)$ & $E$~[MeV] & $P(S{=}\tfrac12)$ & $P(S{=}\tfrac32)$ & $P(l_{QQ})$ & $P(l_{Q\text{-}QQ})$ & $r_{QQ}$ \\
\midrule
$\Omega_{ccc}$ & $0(\tfrac12^-)$ & 5132 & 1.00 & 0.00 & $+0.42$ & $+0.58$ & $0.56$ \\
 & $0(\tfrac32^-)$ & 5133 & 1.00 & 0.00 & $+0.41$ & $+0.59$ & $0.56$ \\
\midrule
$\Omega_{bbb}$ & $0(\tfrac12^-)$ & 14707 & 1.00 & 0.00 & $+0.53$ & $+0.47$ & $0.35$ \\
 & $0(\tfrac32^-)$ & 14707 & 1.00 & 0.00 & $+0.52$ & $+0.48$ & $0.35$ \\
\bottomrule
\end{tabular}
\end{table*}

\begin{table*}[p]
\centering\small\setlength{\tabcolsep}{4pt}
\caption{$1P$ ($L=1$) eigenstates and compositions for the $Q_1Q_1Q_2$ (two identical heavy $+$ one distinct heavy) baryons [$(Q,Q')=(c,b)\,\Omega_{ccb},\ (b,c)\,\Omega_{bbc}$]; the lowest $(3,3,1)$ states are listed. Columns and conventions as in Table~\ref{tab:1P_Qnn}.}
\label{tab:1P_Q1Q1Q2}
\begin{tabular}{ll c cc cccccc}
\toprule
Baryon & $I(J^P)$ & $E$~[MeV] & $P(S{=}\tfrac12)$ & $P(S{=}\tfrac32)$ & $P(l_{QQ})$ & $P(l_{QQ'})$ & $P(l_{Q'\text{-}QQ})$ & $P(l_{Q\text{-}QQ'})$ & $r_{QQ}$ & $r_{QQ'}$ \\
\midrule
$\Omega_{ccb}$ & $0(\tfrac12^-)$ & 8314 & 0.47 & 0.53 & $+0.00$ & $+0.16$ & $+0.25$ & $+0.59$ & $0.46$ & $0.48$ \\
 &  & 8324 & 0.54 & 0.46 & $+0.00$ & $+0.61$ & $+0.66$ & $-0.27$ & $0.46$ & $0.49$ \\
 &  & 8379 & 0.99 & 0.01 & $+0.27$ & $+0.13$ & $+0.00$ & $+0.59$ & $0.60$ & $0.46$ \\
 & $0(\tfrac32^-)$ & 8330 & 0.98 & 0.02 & $+0.00$ & $+0.17$ & $+0.46$ & $+0.36$ & $0.47$ & $0.50$ \\
 &  & 8339 & 0.03 & 0.97 & $+0.00$ & $-0.05$ & $+0.42$ & $+0.63$ & $0.47$ & $0.50$ \\
 &  & 8384 & 0.99 & 0.01 & $+0.27$ & $+0.10$ & $+0.01$ & $+0.62$ & $0.61$ & $0.46$ \\
 & $0(\tfrac52^-)$ & 8351 & 0.00 & 1.00 & $-0.00$ & $+0.20$ & $+0.50$ & $+0.30$ & $0.47$ & $0.51$ \\
\midrule
$\Omega_{bbc}$ & $0(\tfrac12^-)$ & 11509 & 0.98 & 0.02 & $+0.14$ & $-0.07$ & $+0.01$ & $+0.92$ & $0.42$ & $0.40$ \\
 &  & 11552 & 0.16 & 0.84 & $-0.00$ & $+0.57$ & $+0.49$ & $-0.06$ & $0.30$ & $0.49$ \\
 &  & 11580 & 0.86 & 0.14 & $+0.00$ & $+0.49$ & $+0.41$ & $+0.11$ & $0.30$ & $0.51$ \\
 & $0(\tfrac32^-)$ & 11516 & 1.00 & 0.00 & $+0.14$ & $-0.04$ & $+0.00$ & $+0.90$ & $0.42$ & $0.40$ \\
 &  & 11573 & 0.64 & 0.36 & $+0.00$ & $+0.56$ & $+0.39$ & $+0.04$ & $0.30$ & $0.50$ \\
 &  & 11589 & 0.36 & 0.64 & $-0.00$ & $+0.68$ & $+0.34$ & $-0.02$ & $0.30$ & $0.51$ \\
 & $0(\tfrac52^-)$ & 11603 & 0.00 & 1.00 & $+0.00$ & $+0.61$ & $+0.38$ & $+0.02$ & $0.30$ & $0.52$ \\
\bottomrule
\end{tabular}
\end{table*}

\begin{table*}[t]
\centering\small
\caption{$1S$ and $2S$ heavy baryon masses and the radial gap $2S{-}1S$ (MeV). $2S$ is the first radial excitation above the full $1S$ multiplet. Spin-$\tfrac32$ partners are starred. The doubly-heavy and triply-heavy states are pure predictions. 
}
\label{tab:2S}
\begin{tabular}{lccc@{\hskip 1.5em}lccc}
\toprule
Baryon & $1S$ & $2S$ & $2S{-}1S$ & Baryon & $1S$ & $2S$ & $2S{-}1S$\\
\midrule
\multicolumn{8}{l}{\emph{single-charm}}\\
$\Lambda_c$ & 2284 & 2879 & 594 & $\Sigma_c$ & 2442 & 3037 & 595\\
$\Sigma_c^{*}$ & 2528 & 3087 & 559 & $\Xi_c$ & 2472 & 3026 & 554\\
$\Xi_c'$ & 2570 & 3118 & 548 & $\Xi_c^{*}$ & 2654 & 3172 & 517\\
$\Omega_c$ & 2692 & 3240 & 548 & $\Omega_c^{*}$ & 2774 & 3289 & 515\\
\midrule
\multicolumn{8}{l}{\emph{single-bottom}}\\
$\Lambda_b$ & 5614 & 6174 & 560 & $\Sigma_b$ & 5810 & 6356 & 546\\
$\Sigma_b^{*}$ & 5844 & 6375 & 531 & $\Xi_b$ & 5787 & 6302 & 515\\
$\Xi_b'$ & 5922 & 6421 & 499 & $\Xi_b^{*}$ & 5957 & 6442 & 486\\
$\Omega_b$ & 6029 & 6523 & 494 & $\Omega_b^{*}$ & 6064 & 6542 & 478\\
\midrule
\multicolumn{8}{l}{\emph{doubly-heavy $QQq$}}\\
$\Xi_{cc}$ & 3613 & 4070 & 458 & $\Xi_{cc}^{*}$ & 3713 & 4144 & 431\\
$\Omega_{cc}$ & 3728 & 4203 & 475 & $\Omega_{cc}^{*}$ & 3819 & 4268 & 449\\
$\Xi_{bb}$ & 10157 & 10508 & 351 & $\Xi_{bb}^{*}$ & 10200 & 10541 & 342\\
$\Omega_{bb}$ & 10245 & 10614 & 370 & $\Omega_{bb}^{*}$ & 10288 & 10646 & 358\\
\midrule
\multicolumn{8}{l}{\emph{doubly-heavy mixed $bc$}}\\
$\Xi_{bc}$ & 6897 & 7306 & 410 & $\Xi_{bc}'$ & 6944 & 7346 & 402\\
$\Xi_{bc}^{*}$ & 6979 & 7368 & 389 & $\Omega_{bc}$ & 6999 & 7425 & 426\\
$\Omega_{bc}'$ & 7040 & 7460 & 420 & $\Omega_{bc}^{*}$ & 7076 & 7481 & 406\\
\midrule
\multicolumn{8}{l}{\emph{triply-heavy}}\\
$\Omega_{ccb}$ & 8023 & 8465 & 442 & $\Omega_{ccb}^{*}$ & 8059 & 8485 & 426\\
$\Omega_{bbc}$ & 11200 & 11616 & 415 & $\Omega_{bbc}^{*}$ & 11241 & 11643 & 402\\
$\Omega_{ccc}$ & 4833 & 5294 & 461 & $\Omega_{bbb}$ & 14377 & 14804 & 427\\
\bottomrule
\end{tabular}
\end{table*}

\begin{table*}[p]
\centering\small\setlength{\tabcolsep}{4pt}
\caption{$1P$ ($L=1$) eigenstates and compositions for the light baryons with three identical quarks [$q=n$ ($N,\Delta$), $s$ ($\Omega$)]. These states are not included in the fit. The $I=\tfrac12$ nucleon has $(2,2,1)$ states for $J^P=(\tfrac12^-,\tfrac32^-,\tfrac52^-)$. The $\Delta$ ($I=\tfrac32$) and the $\Omega$ ($sss$) each have $(1,1,0)$ states. Their lowest $\tfrac52^-$ level belongs to the $2P$ states. Columns and conventions as in Table~\ref{tab:1P_Qnn}.}
\label{tab:1P_Lqqq}
\begin{tabular}{ll c cc ccc}
\toprule
Baryon & $I(J^P)$ & $E$~[MeV] & $P(S{=}\tfrac12)$ & $P(S{=}\tfrac32)$ & $P(l_{qq})$ & $P(l_{q\text{-}qq})$ & $r_{qq}$ \\
\midrule
$N$ & $\tfrac12(\tfrac12^-)$ & 1380 & 0.78 & 0.22 & $+0.67$ & $+0.33$ & $0.93$ \\
 &  & 1532 & 0.24 & 0.76 & $+0.06$ & $+0.94$ & $1.01$ \\
 & $\tfrac12(\tfrac32^-)$ & 1441 & 1.00 & 0.00 & $+0.33$ & $+0.67$ & $1.02$ \\
 &  & 1626 & 0.00 & 1.00 & $+0.79$ & $+0.21$ & $1.07$ \\
 & $\tfrac12(\tfrac52^-)$ & 1596 & 0.00 & 1.00 & $+0.39$ & $+0.61$ & $1.12$ \\
\midrule
$\Delta$ & $\tfrac32(\tfrac12^-)$ & 1563 & 1.00 & 0.00 & $+0.56$ & $+0.44$ & $1.05$ \\
 & $\tfrac32(\tfrac32^-)$ & 1566 & 1.00 & 0.00 & $+0.55$ & $+0.45$ & $1.05$ \\
\midrule
$\Omega$ & $0(\tfrac12^-)$ & 2012 & 1.00 & 0.00 & $+0.46$ & $+0.54$ & $0.86$ \\
 & $0(\tfrac32^-)$ & 2013 & 1.00 & 0.00 & $+0.45$ & $+0.55$ & $0.86$ \\
\bottomrule
\end{tabular}
\end{table*}

\begin{table*}[p]
\centering\small\setlength{\tabcolsep}{4pt}
\caption{$1P$ ($L=1$) eigenstates and compositions for the light baryons with one identical quark pair [$(q,q')=(n,s)$ ($\Lambda,\Sigma$), $(s,n)$ ($\Xi$)]; the lowest $(3,3,1)$ states are listed. These states are not included in the fit. Columns and conventions as in Table~\ref{tab:1P_Qnn}.}
\label{tab:1P_Lqqp}
\begin{tabular}{ll c cc cccccc}
\toprule
Baryon & $I(J^P)$ & $E$~[MeV] & $P(S{=}\tfrac12)$ & $P(S{=}\tfrac32)$ & $P(l_{qq})$ & $P(l_{qq'})$ & $P(l_{q'\text{-}qq})$ & $P(l_{q\text{-}qq'})$ & $r_{qq}$ & $r_{qq'}$ \\
\midrule
$\Lambda$ & $0(\tfrac12^-)$ & 1470 & 0.99 & 0.01 & $+0.02$ & $+0.25$ & $+0.51$ & $+0.21$ & $0.76$ & $0.90$ \\
 &  & 1593 & 0.75 & 0.25 & $+0.62$ & $-0.02$ & $+0.06$ & $+0.34$ & $0.98$ & $0.82$ \\
 &  & 1714 & 0.27 & 0.73 & $-0.28$ & $+0.60$ & $+0.01$ & $+0.67$ & $1.09$ & $0.88$ \\
 & $0(\tfrac32^-)$ & 1540 & 1.00 & 0.00 & $+0.02$ & $+0.10$ & $+0.60$ & $+0.29$ & $0.82$ & $0.98$ \\
 &  & 1658 & 0.99 & 0.01 & $+0.25$ & $+0.13$ & $+0.09$ & $+0.53$ & $1.14$ & $0.91$ \\
 &  & 1811 & 0.01 & 0.99 & $+0.79$ & $-0.05$ & $-0.00$ & $+0.26$ & $1.18$ & $0.93$ \\
 & $0(\tfrac52^-)$ & 1787 & 0.00 & 1.00 & $+0.28$ & $+0.21$ & $+0.00$ & $+0.51$ & $1.24$ & $0.96$ \\
\midrule
$\Sigma$ & $1(\tfrac12^-)$ & 1604 & 0.70 & 0.30 & $+0.18$ & $+0.07$ & $-0.21$ & $+0.96$ & $0.92$ & $0.88$ \\
 &  & 1707 & 0.33 & 0.67 & $-0.08$ & $+1.16$ & $+0.78$ & $-0.86$ & $0.96$ & $0.97$ \\
 &  & 1733 & 0.97 & 0.03 & $+0.26$ & $+0.33$ & $+0.50$ & $-0.09$ & $1.03$ & $0.97$ \\
 & $1(\tfrac32^-)$ & 1633 & 0.99 & 0.01 & $+0.06$ & $+0.15$ & $+0.31$ & $+0.48$ & $0.98$ & $0.96$ \\
 &  & 1734 & 0.71 & 0.29 & $+0.29$ & $+0.10$ & $+0.27$ & $+0.33$ & $1.06$ & $0.97$ \\
 &  & 1772 & 0.30 & 0.70 & $+0.03$ & $-0.03$ & $-0.04$ & $+1.05$ & $0.96$ & $1.01$ \\
 & $1(\tfrac52^-)$ & 1747 & 0.00 & 1.00 & $+0.00$ & $+0.33$ & $+0.59$ & $+0.09$ & $0.94$ & $1.08$ \\
\midrule
$\Xi$ & $\tfrac12(\tfrac12^-)$ & 1748 & 0.85 & 0.15 & $+0.15$ & $+0.33$ & $+0.23$ & $+0.29$ & $0.84$ & $0.85$ \\
 &  & 1862 & 0.21 & 0.79 & $-0.01$ & $+0.23$ & $+0.26$ & $+0.52$ & $0.74$ & $0.97$ \\
 &  & 1881 & 0.95 & 0.05 & $+0.11$ & $+0.41$ & $-0.02$ & $+0.50$ & $0.85$ & $0.95$ \\
 & $\tfrac12(\tfrac32^-)$ & 1798 & 1.00 & 0.00 & $+0.30$ & $+0.04$ & $+0.03$ & $+0.62$ & $0.95$ & $0.88$ \\
 &  & 1888 & 0.86 & 0.14 & $+0.06$ & $+0.46$ & $+0.29$ & $+0.19$ & $0.80$ & $1.01$ \\
 &  & 1960 & 0.14 & 0.86 & $+0.01$ & $+0.61$ & $+0.70$ & $-0.32$ & $0.76$ & $1.03$ \\
 & $\tfrac12(\tfrac52^-)$ & 1939 & 0.00 & 1.00 & $+0.00$ & $+0.37$ & $+0.39$ & $+0.23$ & $0.77$ & $1.09$ \\
\bottomrule
\end{tabular}
\end{table*}

\begin{table}[htbp]
\centering\small
\caption{Light ($u,d,s$) $1S$ and $2S$ baryon masses and the radial gap $2S{-}1S$ (MeV).
The last two columns give the experimental candidate with the same $J^P$ and the residual $\Delta=M_{2S}^{\rm th}-M_{\rm exp}$, respectively. Each radial excitation mass is predicted to be $\sim200$--$340$~MeV too high. These states are not included in the fit.}
\label{tab:2S_light}
\begin{tabular}{lcccccr}
\toprule
Baryon & $J^P$ & $M_{1S}$ & $M_{2S}$ & $2S{-}1S$ & Exp.\ ($2S$ cand.) & $\Delta$\\
\midrule
$N$ & $\tfrac12^+$ & 926 & 1675 & 749 & $N(1440)$ & $+235$\\
$\Lambda$ & $\tfrac12^+$ & 1120 & 1801 & 681 & $\Lambda(1600)$ & $+201$\\
$\Sigma$ & $\tfrac12^+$ & 1187 & 1887 & 700 & $\Sigma(1660)$ & $+227$\\
$\Xi$ & $\tfrac12^+$ & 1330 & 1976 & 646 & -- & --\\
$\Delta$ & $\tfrac32^+$ & 1219 & 1907 & 689 & $\Delta(1600)$ & $+337$\\
$\Sigma^{*}$ & $\tfrac32^+$ & 1393 & 2026 & 633 & -- & --\\
$\Xi^{*}$ & $\tfrac32^+$ & 1547 & 2140 & 593 & -- & --\\
$\Omega$ & $\tfrac32^+$ & 1687 & 2278 & 592 & -- & --\\
\bottomrule
\end{tabular}
\end{table}

\clearpage
\bibliographystyle{apsrev4-2}
\bibliography{ref}

\end{document}